\documentclass[11pt]{article}
\usepackage[margin=1in]{geometry}
\usepackage{amsmath,amssymb,amsthm}
\usepackage[round,authoryear]{natbib}
\usepackage{hyperref}
\hypersetup{hidelinks}
\usepackage{microtype}
\newtheorem{theorem}{Theorem}
\newtheorem{lemma}{Lemma}
\newtheorem{proposition}{Proposition}
\title{Safety--Performance Frontiers in Data-Driven Monopoly Pricing}
\author{Yiheng Feng\\School of Management, Fudan University\\\texttt{yhfeng25@m.fudan.edu.cn}}
\date{}
\begin{document}
\maketitle
\begin{abstract}
Historical valuation data guide posted-price decisions even when past customers differ from future customers. A seller seeking protection against distribution shift may wish to preserve a benchmark performance guarantee while improving robustness to a broader class of shifts. We study this tradeoff for independent, potentially nonidentically distributed full-valuation observations. Kolmogorov heterogeneity provides the benchmark model, for which empirical revenue maximization has a certified regret guarantee; one-Wasserstein heterogeneity provides the broader stress model at the same radius. We minimize worst-case Wasserstein regret subject to a budget on worst-case Kolmogorov regret for the same, possibly randomized, data-dependent pricing policy. On a fixed H\"older interval-mass class, we characterize the asymptotic minimax safety--performance frontier uniformly over eligible budgets. The frontier moves from a regularity-controlled regime through an inverse-budget regime to the Wasserstein square-root rate. On fixed bounded-density classes with sufficiently large density ceilings, a Kolmogorov budget proportional to the heterogeneity radius imposes a strictly positive gap relative to the Wasserstein-only minimax value at the square-root scale. We further characterize the Kolmogorov budget required for exact Wasserstein-oracle access and establish supporting finite-sample and finite-class results. The findings quantify how protecting benchmark performance constrains robust pricing under broader valuation shifts.
\end{abstract}

\section{Introduction}\label{sec:introduction}

Posted prices are often chosen from past demand, yet a seller's next customer may face economic conditions different from those represented in the data. This discrepancy matters particularly for pricing: moving even a small amount of valuation mass across a posted price can change the purchase probability discontinuously. More historical observations reduce sampling error, but they do not identify how much of an observed sales pattern will persist. The central decision is therefore not simply how to learn a revenue-maximizing price, but how to protect its performance against plausible differences between historical and future valuations.

We study a seller who observes \emph{full valuations} from independent, potentially nonidentically distributed historical customers and selects one data-dependent, possibly randomized pricing policy. Two notions of distributional heterogeneity lead to different performance guarantees. Kolmogorov distance controls the largest discrepancy between distribution functions, while one-Wasserstein distance measures the transportation cost of shifting valuation mass. On $[0,1]$, the Kolmogorov ambiguity model is contained in the Wasserstein model at a common radius. Empirical revenue maximization provides an asymptotic Kolmogorov-regret guarantee of at most $2\varepsilon$. A seller may preserve this certified benchmark while seeking protection against the broader Wasserstein shifts. This leads to a constrained pricing problem: how small can worst-case Wasserstein regret become when the \emph{same policy} must respect a prescribed Kolmogorov-regret budget?

The information structure is important for this question. Each historical marginal must individually lie within the chosen metric radius of the future law. In experiment $F$, historical marginals are otherwise unrestricted; in experiment $A$, they must also belong to the specified future-law class. Both evaluations of a shared policy are made within the same fixed experiment, and they allow arbitrary measurable randomized, sample-size-dependent and order-dependent policies. Section~\ref{sec:model} makes the sequence of expectation, environmental worst case, sample-size limit and policy optimization precise.

\citet{BesbesMaMouchtaki2025} study independent heterogeneous histories and establish metric-specific pricing guarantees: sample average approximation performs well under Kolmogorov heterogeneity, while controlled price reductions attain the square-root Wasserstein rate. These results supply two natural reference points for a seller who wishes to retain a certified benchmark guarantee while preparing for broader shifts. The resulting question concerns the \emph{joint risks} of one pricing policy as its Kolmogorov safety allowance varies. The general principle of controlling baseline risk while adapting to misspecification has precedents in \citet{ArmstrongKlineSun2025} and in robust optimization \citep{BertsimasSim2004}. Related pricing research examines randomized robust policies \citep{BergemannSchlag2011,GuanMisic2026}, moment ambiguity \citep{VanEckKleerVanLeeuwaarden2026}, and competitive ratios under partial distributional information \citep{VanEckKleerVanLeeuwaarden2026MMOR}. Our analysis characterizes the pricing-specific frontier and the cost of its safety constraint under heterogeneous full-valuation observations.

\subsection{Main results and implications}

Our first contribution determines how the allowable Kolmogorov regret governs the best achievable Wasserstein \emph{rate}. On the radius-independent class $\mathcal H_{\alpha,5}$, which satisfies $\mu(I)\le 5|I|^\alpha$ for every interval $I\subseteq[0,1]$ and fixed $0<\alpha<1$, consider a common heterogeneity radius $\varepsilon$, any safety budget $b\ge 2\varepsilon$, and either experiment $E\in\{F,A\}$. The optimal asymptotic Wasserstein regret $G_E(b)$ satisfies
\begin{equation}\label{eq:intro-holder-frontier}
G_E(b)\asymp_{\alpha}
\max\!\left\{\sqrt{\varepsilon},\,
\min\!\left\{\varepsilon^{\alpha/(1+\alpha)},\,
\frac{\varepsilon}{b}\right\}\right\}.
\end{equation}
The uniform comparison holds across eligible budget schedules as $\varepsilon\downarrow0$, after the sample-size limit. It reveals how the safety allowance changes the achievable rate: future-law regularity governs tight budgets, an intermediate range exhibits inverse-budget improvement, and larger allowances attain the square-root Wasserstein rate. Empirical pricing and controlled price reductions establish matching upper bounds; identical historical distributions under competing futures yield the all-policy lower bound. The three regions provide a rate-level characterization of the tradeoff.

The frontier also raises an exact-value question. Achieving the Wasserstein square-root \emph{rate} and attaining the unrestricted Wasserstein-only minimax \emph{value} are different requirements. To examine the latter, let $W_E^\star$ denote that value and $b_E^W$ the least Kolmogorov risk among entire policy sequences attaining it. On the same fixed H\"older class, in both experiments,
\begin{equation}\label{eq:intro-oracle-access}
b_E^W\asymp_{\alpha}\sqrt{\varepsilon}.
\end{equation}
A common-history separation gives the necessary budget for arbitrary randomized, sample-dependent policies. Diagonal selection establishes the existence of exactly Wasserstein-minimax policy sequences and the corresponding upper certificate. This refines the frontier by identifying the safety scale associated with \emph{exact oracle access}.

Our second main result establishes a strict global adaptation gap on a fixed bounded-density class $\mathcal D_B$. For every fixed $B>17/8$ and safety multiplier $L\ge2$, the optimal Wasserstein regret subject to a Kolmogorov budget $L\varepsilon$ exceeds the unrestricted Wasserstein-only minimax value by a positive amount at the $\sqrt\varepsilon$ scale in both experiments. The ratio of the two global minimax values has lower limit strictly above one. A common-history, two-future construction provides the constrained lower certificate, while a separate class-specific policy provides the Wasserstein-only upper certificate. Thus, protecting benchmark performance creates a first-order cost on a fixed distribution class.

Two supporting analyses connect the asymptotic comparisons to policy construction and finite examples. A measurable randomized empirical-price undercutting rule has simultaneous \emph{finite-sample} guarantees for independent non-IID observations. An exactly solvable \emph{fixed three-law class} combines a common-history, multiple-future lower bound with a historical classifier that attains the global statistical upper bound. The former supplies an implementable policy certificate; the latter reveals the geometry of a constrained frontier in a finite class.

Section~\ref{sec:model} introduces the model and performance measures. Section~\ref{sec:holder-and-oracle} establishes the H\"older rate frontier and then exact Wasserstein-oracle access; Section~\ref{sec:density-gap} identifies the bounded-density adaptation gap. Sections~\ref{sec:finite-sample} and \ref{sec:fc-frontier} develop the finite-sample policy and exact fixed-class calculation, respectively. Section~\ref{sec:ur-robustness} returns to the H\"older frontier and shows that its rates persist under \emph{fixed comparable} but unequal Kolmogorov and Wasserstein radii, with regularity imposed on the future law only.

\section{Model and Performance Criteria}\label{sec:model}

\subsection{Posted-price decisions and observations}\label{subsec:model-pricing}
Let $\mathcal P$ denote the Borel probability laws on $[0,1]$, and let $\varnothing\ne\mathcal C\subseteq\mathcal P$ be a specified class of possible future valuation laws. A future customer has valuation $V\sim\mu\in\mathcal C$. A seller posts $p\in[0,1]$, and the customer purchases if and only if $V\ge p$. We include purchase at equality. For a future valuation law, expected revenue and its maximum are
\begin{equation}\label{eq:model-revenue}
r_\mu(p)=p\,\mu([p,1]),\qquad
r_\mu^\star=\max_{p\in[0,1]}r_\mu(p).
\end{equation}
For a randomized price law $q\in\mathcal P$, write $r_\mu(q)=\int r_\mu(p)\,q(dp)$ and define its regret by
\begin{equation}\label{eq:model-loss}
\ell_\mu(q)=r_\mu^\star-r_\mu(q)\ge0.
\end{equation}
The seller then chooses a price from the \emph{full} valuation vector $X_{1:n}=(X_1,\ldots,X_n)$, with independent observations $X_i\sim\nu_i\in\mathcal P$. The unknown marginals $\nu_i$ may differ. Full valuations make $p\mathbf 1\{X_i\ge p\}$ observable for every candidate $p$ and observation, so historical valuations reveal counterfactual revenue at every candidate price, including prices not previously posted.

\subsection{Metric heterogeneity and the two experiments}\label{subsec:model-experiments}
For $\eta\in\mathcal P$ let $F_\eta(x)=\eta([0,x])$. We use the Kolmogorov and one-Wasserstein distances
\begin{equation}\label{eq:model-distances}
\begin{split}
d_K(\eta,\xi)&=\sup_{x\in[0,1]}|F_\eta(x)-F_\xi(x)|,\\
d_W(\eta,\xi)&=W_1(\eta,\xi)
=\inf_{\gamma\in\Gamma(\eta,\xi)}\int|v-w|\,\gamma(dv,dw)
=\int_0^1|F_\eta(x)-F_\xi(x)|\,dx,
\end{split}
\end{equation}
where $\Gamma(\eta,\xi)$ is the set of couplings with marginals $\eta$ and $\xi$. On $[0,1]$, these distances satisfy $d_W\le d_K$. Fix a radius $\varepsilon\ge0$. We require $d(\nu_i,\mu)\le\varepsilon$ \emph{for every} historical marginal $i$; a bound only on the average historical law would describe a different information structure.

For $d\in\{K,W\}$, $n\ge1$, and $E\in\{F,A\}$ define the sets of admissible environments
\begin{equation}\label{eq:model-environments}
\begin{split}
\mathfrak E_{d,n}^{F}(\mathcal C,\varepsilon)
&=\bigl\{(\mu,\nu_{1:n})\in\mathcal C\times\mathcal P^n:
       d(\nu_i,\mu)\le\varepsilon\ \forall i\bigr\},\\
\mathfrak E_{d,n}^{A}(\mathcal C,\varepsilon)
&=\bigl\{(\mu,\nu_{1:n})\in\mathcal C\times\mathcal C^n:
       d(\nu_i,\mu)\le\varepsilon\ \forall i\bigr\}.
\end{split}
\end{equation}
Experiment $F$ allows historical laws outside $\mathcal C$; experiment $A$ restricts each historical marginal to $\mathcal C$. This distinction remains consequential even in $A$ when sample averages are informative: for a nonconvex $\mathcal C$, the barycenter $\bar\nu_n=n^{-1}\sum_{i=1}^n\nu_i$ need not lie in $\mathcal C$. At a common radius, $\mathfrak E_{K,n}^E\subseteq\mathfrak E_{W,n}^E$. We hold the experiment $E$ fixed when comparing $K$ and $W$, changing only the metric that defines admissible heterogeneity.

Two classes used in the principal results are, for fixed $0<\alpha<1$ and $B\ge1$,
\begin{equation}\label{eq:model-classes}
\begin{split}
\mathcal H_{\alpha,5}
&=\bigl\{\mu\in\mathcal P:\ \mu(I)\le5|I|^\alpha
\quad\text{for every interval }I\subseteq[0,1]\bigr\},\\
\mathcal D_B
&=\bigl\{\mu\in\mathcal P:\ \mu(dx)=f(x)\,dx,
\quad0\le f\le B\ \text{a.e.}\bigr\}.
\end{split}
\end{equation}
The H\"older condition controls the probability of short valuation intervals under the \emph{future distribution}, despite the discontinuity of posted-price sales at the threshold. The density class imposes a fixed ceiling. Neither restriction applies to historical marginals in experiment $F$.

\subsection{Policies and worst-case expected regret}\label{subsec:model-risks}
An admissible pricing policy is a sequence $\pi=(\pi_n)_{n\ge1}$ of measurable stochastic kernels
$\pi_n:[0,1]^n\rightsquigarrow[0,1]$.
Given a realized, ordered sample $x_{1:n}$, the kernel $\pi_n(\cdot\mid x_{1:n})$ specifies the distribution of the posted price; point masses represent deterministic decisions. Policies may depend on $n$, $\varepsilon$, $\mathcal C$, and the prescribed performance budget, and need not be invariant to permutations of the observations. For the fixed experiment $E$, the \emph{same} policy is evaluated under $K$ and $W$; it is not told which metric's admissibility model governs the data, nor required to select between experiments $F$ and $A$.

For $d\in\{K,W\}$ and $E\in\{F,A\}$, define finite-sample worst-case expected regret by
\begin{equation}\label{eq:model-finite-risk}
\begin{split}
R_{d,n}^{E}(\pi;\mathcal C,\varepsilon)
=\sup_{(\mu,\nu_{1:n})\in\mathfrak E_{d,n}^{E}(\mathcal C,\varepsilon)}
\int_{[0,1]^n}\ell_\mu\!\left(\pi_n(\cdot\mid x_{1:n})\right)
\prod_{i=1}^n\nu_i(dx_i).
\end{split}
\end{equation}
For each environment, the integral averages over the independent observations, and $\ell_\mu$ incorporates the price randomization. We then maximize this expected regret over admissible environments. The asymptotic risk of the policy sequence is
\begin{equation}\label{eq:model-asymptotic-risk}
R_d^{E}(\pi;\mathcal C,\varepsilon)
=\limsup_{n\to\infty}R_{d,n}^{E}(\pi;\mathcal C,\varepsilon).
\end{equation}
The order matters: at each fixed $\varepsilon$, the finite-$n$ environmental supremum precedes the sample-size $\limsup$, and optimization over policy sequences precedes any small-radius limit $\varepsilon\downarrow0$. This ordering retains sampling uncertainty in the statistical risks defined by \eqref{eq:model-finite-risk}. The nesting of metric balls also gives
\begin{equation}\label{eq:model-risk-order}
R_K^E(\pi;\mathcal C,\varepsilon)
\le R_W^E(\pi;\mathcal C,\varepsilon)
\end{equation}
for every admissible $\pi$.

\subsection{Safety budgets, frontier, and oracle benchmark}\label{subsec:model-frontier}
Let $\Pi$ be the class of all such policy sequences. For a Kolmogorov safety budget $b\ge0$, define the constrained Wasserstein frontier and the unconstrained Wasserstein-only oracle by
\begin{equation}\label{eq:model-frontier}
\begin{split}
G_E(b;\mathcal C,\varepsilon)
&=\inf_{\pi\in\Pi:\ R_K^E(\pi;\mathcal C,\varepsilon)\le b}
        R_W^E(\pi;\mathcal C,\varepsilon),\\
W_E^\star(\mathcal C,\varepsilon)
&=\inf_{\pi\in\Pi}R_W^E(\pi;\mathcal C,\varepsilon),
\end{split}
\end{equation}
with the infimum of an empty feasible set interpreted as $+\infty$. For a fixed safety multiplier $L\ge2$, set
\begin{equation}\label{eq:model-adaptation-gap}
A_L^E(\mathcal C,\varepsilon)=G_E(L\varepsilon;\mathcal C,\varepsilon),
\qquad
\Delta_L^E(\mathcal C,\varepsilon)=A_L^E(\mathcal C,\varepsilon)-W_E^\star(\mathcal C,\varepsilon).
\end{equation}
The exact-oracle access budget measures the least Kolmogorov risk among asymptotic policy sequences attaining the exact Wasserstein-only value:
\begin{equation}\label{eq:model-access-budget}
b_E^W(\mathcal C,\varepsilon)=
\min_{\substack{\pi\in\Pi:\\R_W^E(\pi;\mathcal C,\varepsilon)=W_E^\star(\mathcal C,\varepsilon)}}
R_K^E(\pi;\mathcal C,\varepsilon).
\end{equation}
This is a \emph{minimum} at the policy-sequence level: stagewise diagonal selection first attains the Wasserstein-risk infimum and then the least Kolmogorov risk among such sequences. No finite-sample minimizer is required. As a function of $b$, $G_E$ is nonincreasing and convex, the latter following by randomizing between entire policy sequences. The empirical-revenue-maximizing policy has $R_K^E\le2\varepsilon$ in either experiment, so the safety-constrained problem is feasible for $b\ge2\varepsilon$. In all small-radius claims below, $b=b(\varepsilon)$ may vary and the order is fixed: expectation, environmental supremum, $n\to\infty$, optimization in \eqref{eq:model-frontier}, and then $\varepsilon\downarrow0$.

The tradeoff can already be seen in a pair of worlds with identical historical observations but different optimal future prices. Let $\delta_x$ be a point mass at $x$. Fix $0<\varepsilon<1/16$ and set $m=2\sqrt\varepsilon$ and $h=\sqrt\varepsilon/2$. Take the common historical marginal $\nu$ and two future laws
\[
\begin{aligned}
\nu=\mu_K&=\frac{1-m}{2}\,\delta_{1/2}+\frac{1+m}{2}\,\delta_1,\\
\mu_W&=\frac{1-m}{2}\,\delta_{1/2}+m\,\delta_{1-h}
          +\frac{1-m}{2}\,\delta_1.
\end{aligned}
\]
The illustrative future class $\mathcal C_\varepsilon=\{\mu_K,\mu_W\}$ \emph{depends on the radius} $\varepsilon$. The identical i.i.d. history with law $\nu$ is admissible under $K$ for $\mu_K$ and under $W$ for $\mu_W$, in \emph{both} experiments: $d_K(\nu,\mu_K)=0$, while transporting mass $m$ from $1$ to $1-h$ gives $d_W(\nu,\mu_W)=mh=\varepsilon$. Their respective optimal prices are $1$ and $1-h$, and the cross-regrets are exactly
\begin{equation}\label{eq:model-conflict}
\ell_{\mu_K}(\delta_{1-h})=\frac{h(1+m)}{2},\qquad
\ell_{\mu_W}(\delta_1)=m-\frac{h(1+m)}{2}.
\end{equation}
Both cross-regrets have order $\sqrt\varepsilon$. Thus a discount that protects the Wasserstein-compatible future can be costly relative to an $O(\varepsilon)$ Kolmogorov safety budget. Because the illustrative atomic class $\mathcal C_\varepsilon$ varies with the radius, this example captures a \emph{local} conflict. The subsequent theorems establish all-policy bounds on the fixed classes in \eqref{eq:model-classes}.

\section{The H\"older Frontier and Exact Wasserstein-Oracle Access}\label{sec:holder-and-oracle}

\subsection{The Uniform H\"older Safety--Performance Frontier}\label{sec:holder-frontier}

The opening question is how the Kolmogorov safety budget limits a policy's ability to protect against Wasserstein shifts. Such a shift can move valuation mass across a posted price at small transportation cost, while changing purchases discontinuously. The amount of mass near a price threshold consequently governs the loss. We impose regularity on future valuations and study how the downward price adjustments of \citet{BesbesMaMouchtaki2025} interact with a Kolmogorov-regret requirement.

Fix $0<\alpha<1$ and the radius-independent class $\mathcal H_{\alpha,5}$ in \eqref{eq:model-classes}: every interval of length $u$ has future probability at most $5u^\alpha$. Its distribution functions obey $|F_\mu(x)-F_\mu(y)|\le5|x-y|^\alpha$. More importantly, for $\mu\in\mathcal H_{\alpha,5}$ and an \emph{arbitrary} law $\nu$, the interval-mass restriction and the integrated-CDF formula for $W_1$ imply
\begin{equation}\label{eq:holder-interpolation}
 d_K(\mu,\nu)\le C_\alpha\,d_W(\mu,\nu)^{\alpha/(1+\alpha)}.
\end{equation}
Because the bound uses regularity only of the future law, it applies even when historical laws are not H\"older in experiment $F$. Consequently, sample average approximation (SAA) is Kolmogorov-safe at risk $2\varepsilon$ and has Wasserstein regret at most of order $\varepsilon^{\alpha/(1+\alpha)}$. Lowering its price can improve that Wasserstein rate toward the square-root scale, but incurs an additional Kolmogorov safety cost. The frontier quantifies this exchange.

\begin{theorem}[Uniform H\"older safety--performance frontier]\label{thm:holder-frontier}
For each fixed $0<\alpha<1$, there exist $\varepsilon_0(\alpha)>0$ and constants $0<c_\alpha\le C_\alpha<\infty$ such that, for both $E=F$ and $E=A$, all $0<\varepsilon\le\varepsilon_0(\alpha)$, and \emph{every} $b\ge2\varepsilon$,
\begin{equation}\label{eq:holder-frontier}
 c_\alpha\,\Phi_\alpha(\varepsilon,b)
 \le G_E(b;\mathcal H_{\alpha,5},\varepsilon)
 \le C_\alpha\,\Phi_\alpha(\varepsilon,b),\qquad
 \Phi_\alpha(\varepsilon,b)=
 \max\!\left\{\sqrt\varepsilon,
 \min\!\left\{\varepsilon^{\alpha/(1+\alpha)},\frac{\varepsilon}{b}\right\}\right\}.
\end{equation}
The constants do not depend on $b$, even for a budget schedule $b=b(\varepsilon)$. The risks are the worst-case expected, arbitrary-policy, potentially non-IID asymptotic risks of \eqref{eq:model-finite-risk}--\eqref{eq:model-frontier}; the sample-size limit is taken at each fixed $\varepsilon$ before the small-radius statement.
\end{theorem}

Theorem~\ref{thm:holder-frontier} identifies three \emph{rate} regions. For $2\varepsilon\le b\lesssim_\alpha\varepsilon^{1/(1+\alpha)}$, future-law regularity controls regret at $\varepsilon^{\alpha/(1+\alpha)}$. When $\varepsilon^{1/(1+\alpha)}\lesssim_\alpha b\lesssim\sqrt\varepsilon$, additional safety allowance reduces regret at rate $\varepsilon/b$. Once $b\gtrsim\sqrt\varepsilon$, the Wasserstein-only rate $\sqrt\varepsilon$ is available. These are constant-factor regions, not assertions about exact kinks or leading coefficients.

The upper-bound mechanism makes the tradeoff explicit. All policies start from the same empirical full-valuation revenue maximizer $\widehat p_n$ under both metrics. SAA posts $\widehat p_n$ and remains feasible even at $b=2\varepsilon$. For a width $t>0$, deterministic undercutting posts $(\widehat p_n-t)_+$; its asymptotic risks satisfy $R_K^E\le2\varepsilon+t$ and $R_W^E\le2\sqrt\varepsilon+t+\varepsilon/t$. Taking $t=b-2\varepsilon$ when $b$ is at least the H\"older width gives the intermediate regime. Alternatively, draw $U$ uniformly from $[0,t]$ and post $(\widehat p_n-U)_+$. Smoothing the sales threshold improves the bounds to $R_K^E\le2\varepsilon+t/2$ and $R_W^E\le2\sqrt\varepsilon+t/2+\varepsilon/t$; $t=\sqrt{2\varepsilon}$ is feasible at $b\ge\sqrt\varepsilon$ for small $\varepsilon$. Uniform non-IID empirical deviations establish the guarantees for these policies under heterogeneous sampling. Their proofs, including the positive-part boundary, appear in Appendix~\ref{app:holder-proof}.

To show that no other policy can improve the three rates, we construct futures with the \emph{same} historical observations. Let $a=1/2$, $h=(4^{-\alpha}\varepsilon)^{1/(1+\alpha)}$, and $\delta=(4h)^\alpha$, so $\delta h=\varepsilon$. A historical law $\nu$ places mass $\delta$ uniformly just above $a$ and its remaining mass farther to the right. The competing Wasserstein future $\mu_m$ shifts only $m\le\delta/4$ of this first mass to just below $a$, over distance $D=\varepsilon/m$; the shifted block has width $s=mh/\delta$. Both futures obey the every-interval H\"older condition, and $W_1(\mu_m,\nu)=\varepsilon$. With $Q=am-D-s$, $D>s$, and $\lambda=\max\{1,Q/(D-s)\}$, the price-by-price certificate is
\begin{equation}\label{eq:holder-witness-certificate}
 \ell_{\mu_m}(\delta_p)+\lambda\ell_\nu(\delta_p)\ge Q
 \quad\text{for all }p\in[0,1].
\end{equation}
Because every historical sample has law $\nu^{\otimes n}$ in both worlds, this inequality applies after arbitrary data dependence and price randomization. Using $m=\delta/4$ at low budgets, $m=\varepsilon/(4b)$ within its valid intermediate range, and $m=8\sqrt\varepsilon$ for the universal Wasserstein floor yields the matching three lower rates. The witness lies in both experiments, including the class-restricted experiment $A$.

The rate comparison leaves a further decision question. Even if some safe policy has Wasserstein regret of square-root order, it may not attain the exact unconstrained minimax \emph{value} $W_E^\star$. The exact-value analysis below identifies the Kolmogorov budget required for that equality.

\subsection{Exact Wasserstein-Oracle Access}\label{sec:exact-oracle-access}

The relevant notion of access is the minimum Kolmogorov risk paid by a \emph{whole policy sequence} attaining Wasserstein-only optimality. Precisely, \eqref{eq:model-access-budget} defines $b_E^W(\mathcal C,\varepsilon)$ as the minimum $R_K^E$ among sequences whose $R_W^E$ equals $W_E^\star(\mathcal C,\varepsilon)$. A diagonal-selection argument establishes both minimax attainment and attainment of this minimum in the asymptotic sequence class (Appendix~\ref{app:oracle-attainment}); no finite-sample minimizer is needed. It follows that $G_E(b_E^W;\mathcal C,\varepsilon)=W_E^\star(\mathcal C,\varepsilon)$.

\begin{theorem}[Exact Wasserstein-oracle access on the H\"older class]\label{thm:exact-oracle-access}
Fix $0<\alpha<1$ and $\mathcal C=\mathcal H_{\alpha,5}$. There exists $\varepsilon_0(\alpha)>0$ such that, for both experiments $E\in\{F,A\}$ and all $0<\varepsilon\le\varepsilon_0(\alpha)$,
\begin{equation}\label{eq:exact-oracle-access-bounds}
 \frac{61/16-\sqrt2}{64}\,\sqrt\varepsilon
 \;\le\; b_E^W(\mathcal H_{\alpha,5},\varepsilon)
 \;\le\;\sqrt{2\varepsilon}.
\end{equation}
Consequently,
\begin{equation}\label{eq:exact-oracle-access-rate}
 b_E^W(\mathcal H_{\alpha,5},\varepsilon)=\Theta_\alpha(\sqrt\varepsilon),
 \qquad E=F,A.
\end{equation}
The bounds refer to equality with the exact \emph{asymptotic} minimax value at each fixed radius, with the sample-size limit taken before $\varepsilon\downarrow0$; they do not assert a sharp leading coefficient.
\end{theorem}

An all-policy separation establishes the lower access budget. The high-budget specialization of the common-history certificate in Lemma~\ref{lem:holder-all-prices} gives, for every admissible sequence $\pi$ and either experiment,
\begin{equation}\label{eq:exact-oracle-separation}
 R_W^E(\pi;\mathcal H_{\alpha,5},\varepsilon)
 +64R_K^E(\pi;\mathcal H_{\alpha,5},\varepsilon)
 \ge \frac{61}{16}\sqrt\varepsilon.
\end{equation}
The underlying Wasserstein future transports a selected block of valuation mass across a candidate price, whereas the Kolmogorov future and both histories retain the original law. Because these worlds generate identical observations and the inequality holds at \emph{every} price, randomizing or using sample labels cannot circumvent it. On the other side, a Wasserstein-aware smoothed-price policy provides the universal benchmark
\begin{equation}\label{eq:exact-oracle-universal-benchmark}
 W_E^\star(\mathcal C,\varepsilon)\le\sqrt{2\varepsilon}
 \qquad(E=F,A),
\end{equation}
for any nonempty future class $\mathcal C$ and sufficiently small $\varepsilon$. The construction and its nonidentical-sample bound are proved in Appendix~\ref{app:oracle-universal}.

For an exact Wasserstein-minimax sequence, substituting $R_W^E=W_E^\star$ into \eqref{eq:exact-oracle-separation} and using \eqref{eq:exact-oracle-universal-benchmark} forces the lower budget in \eqref{eq:exact-oracle-access-bounds}. Conversely, exact Wasserstein-minimax sequences exist, and the inclusion of Kolmogorov environments in Wasserstein environments implies $R_K^E\le R_W^E=W_E^\star\le\sqrt{2\varepsilon}$ for any such sequence. Thus a Kolmogorov allowance on the square-root scale suffices for \emph{exact} oracle access. The decisive step is the existence of a Wasserstein-minimax sequence, not merely a construction with the correct rate. This separates access to an exact asymptotic value from finite-sample optimality or the determination of sharp transition coefficients.

\section{A Strict Adaptation Cost Under Bounded Density}\label{sec:density-gap}

The access theorem identifies when the unconstrained Wasserstein minimax value can be reached on the H\"older class. A different issue is whether imposing a tight Kolmogorov budget has a first-order cost: on fixed bounded-density classes, a constrained lower certificate exceeds an unrestricted Wasserstein-only upper certificate at the square-root scale. Bounded density limits the mass that can be transported across a price, making this question especially pointed. We compare the global constrained value $A_L^E(\mathcal D_B,\varepsilon)$ and the Wasserstein-only minimax value $W_E^\star(\mathcal D_B,\varepsilon)$ on the same fixed bounded-density class.

Consider the fixed class $\mathcal D_B$ in \eqref{eq:model-classes}, with density at most $B$; experiment $F$ allows arbitrary historical marginals, whereas experiment $A$ requires them also to lie in $\mathcal D_B$. For fixed $B>1$, define
\begin{equation}\label{eq:density-certificates}
 P_B=\frac{1+\sqrt{1-B^{-2}}}{2},\qquad
 C_B=\sqrt B\,P_B,\qquad
 U_B=\sqrt{2P_B}.
\end{equation}
The two constants play different roles: $C_B$ lower-bounds globally optimal Kolmogorov-safe regret, whereas $U_B$ upper-bounds the globally optimal Wasserstein-only benchmark. A strict separation between these bounds certifies an adaptation cost.

\begin{theorem}[Strict global adaptation gap under bounded density]\label{thm:density-gap}
Fix $B>17/8$, $L\ge2$, and $E\in\{F,A\}$. In the independent, possibly nonidentically distributed full-valuation experiment of Section~\ref{sec:model}, with the sample-size asymptotic taken at each fixed radius before $\varepsilon\downarrow0$,
\begin{equation}\label{eq:density-global-certificates}
 \liminf_{\varepsilon\downarrow0}
 \frac{A_L^E(\mathcal D_B,\varepsilon)}{\sqrt\varepsilon}\ge C_B,
 \qquad
 \limsup_{\varepsilon\downarrow0}
 \frac{W_E^\star(\mathcal D_B,\varepsilon)}{\sqrt\varepsilon}\le U_B.
\end{equation}
Moreover, $W_E^\star(\mathcal D_B,\varepsilon)>0$ for all sufficiently small $\varepsilon>0$, and
\begin{align}
 \liminf_{\varepsilon\downarrow0}
 \frac{\Delta_L^E(\mathcal D_B,\varepsilon)}{\sqrt\varepsilon}
 &\ge C_B-U_B>0,\label{eq:density-strict-difference}\\
 \liminf_{\varepsilon\downarrow0}
 \frac{A_L^E(\mathcal D_B,\varepsilon)}{W_E^\star(\mathcal D_B,\varepsilon)}
 &\ge\frac{C_B}{U_B}>1.\label{eq:density-strict-ratio}
\end{align}
These inequalities concern global asymptotic statistical minimax values and allow arbitrary randomized, sample-dependent pricing policies.
\end{theorem}

The lower-bound construction places density $B$ on two separated intervals, creating nearly tied revenue maxima at distant prices. Kolmogorov safety then favors the slightly better high price, while a Wasserstein perturbation transports a saturated mass block at cost exactly $\varepsilon$ and incurs revenue loss of order $\sqrt\varepsilon$. The identical histories extend the lower certificate to every policy in the global constrained problem.

For the global oracle, every law in $\mathcal D_B$ has an optimal price at most $P_B$. Clipping and rescaling valuations transforms the Wasserstein radius to $\varepsilon/P_B$ while scaling regret by $P_B$. Applying the universal randomized smoothing lemma in Appendix~\ref{app:oracle-universal} to the transformed problem yields $U_B$, sharper than the unrestricted $\sqrt2$ coefficient. Appendix~\ref{app:density-gap} proves the density-specific refinements and statistical limit.

The certificate coefficients cross at $B=17/8$, giving a sufficient condition for strict separation when $B>17/8$. For example, at $B=12$ the certified lower and upper coefficients are approximately $3.458077$ and $1.412983$, giving a difference certificate of $2.045094$ and a ratio certificate of $2.447359$. These values quantify the certified separation between the global minimax values. The global $\mathcal D_2$ adaptation boundary remains unresolved. The next section turns from a global impossibility certificate to the separate, constructive question of what a specified pricing policy guarantees with finitely many heterogeneous observations.

\section{Constructive Finite-Sample Policies}\label{sec:finite-sample}

The preceding theorems characterize asymptotic minimax comparisons. To connect their policy mechanisms with observed data, we now give a computable rule for finitely many independent, heterogeneous valuations. Its simultaneous finite-sample certificates quantify both metric uncertainty and estimation error for the same randomized price. These certificates describe the finite-sample performance of the specified policy.

Given independent, potentially nonidentically distributed full valuations $X_{1:n}$, let
\begin{equation}\label{eq:fs-empirical-rule}
 \begin{aligned}
 \widehat r_n(p)&=\frac{p}{n}\sum_{i=1}^n\mathbf 1\{X_i\ge p\},\qquad
 \widehat p_n=\min\arg\max_{p\in\{0,X_1,\ldots,X_n\}}\widehat r_n(p),\\
 P_{n,t}&=(\widehat p_n-U)_+,\qquad U\sim\operatorname{Unif}[0,t].
 \end{aligned}
\end{equation}
where $t>0$ is prescribed and $U$ is independent of the observations. The deterministic minimum resolves empirical revenue ties, and sorting the valuations implements the procedure in $O(n\log n)$ time. Random undercutting spreads the effective sales threshold over an interval. The resulting kernel is the same in the Kolmogorov and Wasserstein evaluations; neither the historical marginals nor the operative metric are supplied to it.

\begin{proposition}[Finite-sample simultaneous safety and performance]\label{prop:fs-simultaneous}
For every nonempty future class $\mathcal C\subseteq\mathcal P$, $E\in\{F,A\}$, $n\ge1$, $\varepsilon\ge0$, and $t>0$, the policy \eqref{eq:fs-empirical-rule} is a measurable pricing kernel and satisfies
\begin{equation}\label{eq:fs-main-bounds}
 \boxed{\begin{aligned}
 R_{K,n}^E(P_{n,t};\mathcal C,\varepsilon)&\le 2\varepsilon+t+\frac8{\sqrt n},\\
 R_{W,n}^E(P_{n,t};\mathcal C,\varepsilon)&\le 2\sqrt\varepsilon+t+\frac{\varepsilon}{t}+\frac8{\sqrt n}.
 \end{aligned}}
\end{equation}
The risks are the \emph{worst-case expected regrets} of \eqref{eq:model-finite-risk}, including averaging over both the historical sample and the policy's independent randomization.
\end{proposition}

For a prescribed Kolmogorov budget $b$, the stated certificate admits a strictly positive width only if its \emph{certified slack} $s_n=b-2\varepsilon-8/\sqrt n$ is positive. When $\varepsilon>0$ and $s_n>0$, the width
\begin{equation}\label{eq:fs-width-choice}
 t_n^{\mathrm{cert}}=\min\!\left\{\sqrt\varepsilon,\ b-2\varepsilon-\frac8{\sqrt n}\right\}
\end{equation}
minimizes $t+\varepsilon/t$ among positive widths certified by the first inequality. Thus $b>2\varepsilon+8/\sqrt n$ suffices, or equivalently $n>64/(b-2\varepsilon)^2$ when $b>2\varepsilon$. If the slack is nonpositive, \eqref{eq:fs-main-bounds} cannot certify any positive-width policy at that budget; this does \emph{not} prove that no other policy is feasible. The term $8/\sqrt n$ accounts for estimation from independent nonidentical samples; the width $t$ trades additional Kolmogorov regret against robustness to transported valuations. Appendix~\ref{app:finite-sample} proves the bounds, their conditional and high-probability counterparts, and a class-qualified finite-sample statistical lower floor.

The policy connects to the H\"older frontier in Theorem~\ref{thm:holder-frontier}: once $n$ is sufficiently large, the admissible width tracks safety slack; $\varepsilon/t$ expresses the inverse-budget tradeoff, while $t\asymp\sqrt\varepsilon$ yields a square-root upper rate. At the closed low-budget boundary, SAA and future-law regularity remain essential, as in Subsection~\ref{sec:holder-frontier}. This proposition neither replaces the uniform three-regime theorem nor fixes finite-$n$ minimax transition budgets. The next, distinct supporting result instead fixes a three-law class and derives an exact \emph{asymptotic global} frontier, requiring both an information-based lower bound and a separate non-IID statistical upper construction.

\section{Exact Finite-Class Geometry and Statistical Implementation}\label{sec:fc-frontier}

The finite-sample construction offers a broadly applicable policy bound. An exact frontier, however, must account for all futures compatible with the same observed history and for uniform implementation under heterogeneous sampling. We isolate those two tasks on a fixed three-law class: a common-history linear program constrains every policy, and a separate empirical classifier realizes its optimum over all admissible independent non-IID arrays. Unlike the rate frontier in Theorem~\ref{thm:holder-frontier}, the resulting fixed-class calculation has exact breakpoints and shadow prices.

\paragraph{Historical information.}
For $d\in\{K,W\}$, write $B_d(\mu,\varepsilon)=\{\nu\in\mathcal P:d(\nu,\mu)\le\varepsilon\}$. In experiment $F$, every possible average historical distribution for future $\mu$ lies in this ball, and every point of the ball is realized by homogeneous histories. In experiment $A$, the closure of the attainable historical averages is instead
\begin{equation}\label{eq:fc-fibers}
 \mathcal D_d^A(\mu)=\overline{\operatorname{co}}\!\left(\mathcal C\cap B_d(\mu,\varepsilon)\right),
 \qquad \mathcal D_d^F(\mu)=B_d(\mu,\varepsilon).
\end{equation}
The intersection precedes convexification: each historical marginal must be individually admissible in $A$, although its average need not be in $\mathcal C$. The average thereby describes which historical centers are attainable, but not every aspect of a labeled product experiment. The common-center game yields a lower bound; achieving the global statistical frontier also requires a policy valid uniformly over independent, potentially nonidentical arrays. A classifier attains the global upper bound while accommodating the discontinuous revenue functions of the atomic class.

\paragraph{Three future laws.}
Fix $\varepsilon=1/100$ and $\mathcal C_\mathrm{fc}=\{H,U,V\}$ with
\begin{equation}\label{eq:fc-laws}
 \begin{aligned}
 H&=\tfrac14\delta_{1/2}+\tfrac34\delta_{97/150},\\
 U&=\tfrac14\delta_{23/50}+\tfrac34\delta_{97/150},\\
 V&=\tfrac14\delta_{1/2}+\tfrac34\delta_{19/30}.
 \end{aligned}
\end{equation}
The metric geometry is
\begin{equation}\label{eq:fc-metric-table}
\begin{array}{c|ccc}
 & (H,U)&(H,V)&(U,V)\\ \hline
 d_K &1/4&3/4&3/4\\
 d_W &1/100&1/100&1/50
\end{array}.
\end{equation}
In both experiments, history $H$ is Kolmogorov-compatible with future $H$ and Wasserstein-compatible with each of $H,U,V$. It thus forces one pricing rule to protect $H$ under the safety criterion while accommodating two opposing Wasserstein futures. The regret vectors of all prices are componentwise dominated by those at $p_1=1/2$, $p_2=19/30$, or $p_3=97/150$; the exact reduced matrix, including purchase at equality, is
\begin{equation}\label{eq:fc-regret-matrix}
\begin{array}{c|ccc}
 &p_1=1/2&p_2=19/30&p_3=97/150\\ \hline
 \ell_H &0&1/40&3/200\\
 \ell_U &11/100&1/100&0\\
 \ell_V &0&1/40&1/2
\end{array}.
\end{equation}

\begin{theorem}[Exact three-law statistical frontier]\label{thm:fc-exact}
For the fixed class \eqref{eq:fc-laws}, radius $\varepsilon=1/100$, and either experiment $E\in\{F,A\}$, the global asymptotic frontier defined in \eqref{eq:model-frontier} is
\begin{equation}\label{eq:fc-piecewise}
 \boxed{G_E(b;\mathcal C_\mathrm{fc},1/100)=
 \begin{cases}
 +\infty,&b<0,\\[6pt]
 \dfrac{11}{100}-\dfrac{22}{3}b,&0\le b\le\dfrac{33}{12200},\\[7pt]
 \dfrac{1067}{10700}-\dfrac{378}{107}b,&\dfrac{33}{12200}\le b\le\dfrac{11}{500},\\[7pt]
 \dfrac{11}{500},&b\ge\dfrac{11}{500}.
 \end{cases}}
\end{equation}
The Wasserstein-only oracle and its least Kolmogorov-access budget are both
\begin{equation}\label{eq:fc-oracle-access}
 W_E^\star(\mathcal C_\mathrm{fc},1/100)=b_E^W(\mathcal C_\mathrm{fc},1/100)=\frac{11}{500}.
\end{equation}
These equalities hold over all measurable randomized, sample-size-dependent and order-dependent policies, for independent nonidentically distributed histories in both $F$ and $A$.
\end{theorem}

At the safety budget $b=1/50$, a global policy uses, in the shared-history region, the price mixture
\begin{equation}\label{eq:fc-distinguished-policy}
 q_{1/50}=\frac{103}{535}\delta_{1/2}
 +\frac{422}{535}\delta_{19/30}
 +\frac{2}{107}\delta_{97/150},\qquad
 (\ell_H,\ell_U,\ell_V)(q_{1/50})
 =\left(\frac1{50},\frac{311}{10700},\frac{311}{10700}\right).
\end{equation}
The mixture exhausts the Kolmogorov safety budget through its first regret component, while both competing Wasserstein futures are binding. The global constrained cost strictly exceeds the global Wasserstein-only oracle:
\begin{equation}\label{eq:fc-gap}
 G_E(1/50)=\frac{311}{10700},\qquad
 G_E(1/50)-W_E^\star=\frac{189}{26750}>0.
\end{equation}
The supporting Wasserstein weights at this budget are $(97/107,10/107)$ on $(U,V)$, and the Kolmogorov shadow price is $378/107$. For smaller budgets the first affine segment is supported by $U$ alone, with safety multiplier $22/3$; at the oracle plateau the safety multiplier becomes zero. The piecewise slopes thus reveal how the jointly active futures change as the safety budget grows.

To realize the frontier statistically, the policy assigns the common-history mixture near $H$ and treats separated historical arrays differently. An empirical Kolmogorov classifier detects histories sufficiently close to $U$ or $V$ and selects branch-specific prices; a fixed separation inequality rules out the opposite Wasserstein future on each branch. The uniform non-IID concentration bound and exact all-price primal--dual certificates are established in Appendix~\ref{app:fc-proof}. In particular, the center-$H$ linear program gives a lower bound, and the classifier independently establishes that this lower bound is the \emph{global} statistical frontier. At $b=1/50$ the strongest single-K/single-W pair certificate is only $27/2200$, strictly below $311/10700$; subtracting the global oracle from that pairwise quantity would miss the positive adaptation excess.

The atomic calculation demonstrates why a single pairwise comparison may not capture the cost of accommodating several Wasserstein futures, while its exact rational breakpoints are specific to the fixed class. We next return to the nonatomic H\"older class of Subsection~\ref{sec:holder-frontier} and ask whether its rate frontier is stable when the two metric radii differ by fixed factors.

\section{Robustness to Comparable Unequal Radii}\label{sec:ur-robustness}

The final question concerns the robustness of the natural-class rate comparison itself. Kolmogorov safety and Wasserstein performance need not be assessed at identical uncertainty radii. We return to the fixed H\"older class and Theorem~\ref{thm:holder-frontier} to determine whether the three rate regions persist for \emph{different but comparable} radii. This extension examines the stability of the frontier under comparable uncertainty radii.

Write $\varepsilon_K>0$ and $\varepsilon_W>0$ for their respective, independently specified radii. Retaining the experiments and policy class of Section~\ref{sec:model}, define
\begin{equation}\label{eq:ur-definition}
 G_E(b;\mathcal C,\varepsilon_K,\varepsilon_W)
 =\inf_{\pi\in\Pi:\,R_K^E(\pi;\mathcal C,\varepsilon_K)\le b}
 R_W^E(\pi;\mathcal C,\varepsilon_W),\qquad E\in\{F,A\}.
\end{equation}
The seller still commits to the same observation-based policy, evaluated against two possibly different sets of admissible histories. At each fixed pair of radii, expectations precede environmental suprema and then the sample-size $\limsup$.

\begin{proposition}[Comparable unequal-radius H\"older frontier]\label{prop:ur-frontier}
Fix $0<\alpha<1$ and $0<c\le C<\infty$. There exist $\varepsilon_0(\alpha,c,C)>0$ and $0<k_{\alpha,c,C}\le K_{\alpha,c,C}<\infty$ such that, for both $E=F,A$, whenever
\[
0<\varepsilon_W\le\varepsilon_0,\qquad
c\le\frac{\varepsilon_K}{\varepsilon_W}\le C,\qquad
b\ge2\varepsilon_K,
\]
we have, \emph{uniformly} over these radii and budgets,
\begin{equation}\label{eq:ur-frontier}
 k_{\alpha,c,C}\,\Phi_\alpha(\varepsilon_W,b)
 \le G_E(b;\mathcal H_{\alpha,5},\varepsilon_K,\varepsilon_W)
 \le K_{\alpha,c,C}\,\Phi_\alpha(\varepsilon_W,b).
\end{equation}
Here $\Phi_\alpha$ is the rate envelope defined in \eqref{eq:holder-frontier}. The constants are independent of any joint budget/radius schedule satisfying these conditions.
\end{proposition}

The result preserves the same three economic regimes: regularity controls tight budgets, additional safety produces inverse-budget improvement, and Wasserstein robustness imposes a square-root floor. The corresponding rate-transition budgets remain of orders $\varepsilon_W^{1/(1+\alpha)}$ and $\sqrt{\varepsilon_W}$ under bounded radius comparability. The proof uses the equal-radius frontier by monotonicity except when $\varepsilon_K<\varepsilon_W$ and $2\varepsilon_K\le b<2\varepsilon_W$. In that interval a measurable empirical-revenue-maximizing price is safe \emph{at} the limiting budget $2\varepsilon_K$ and has Wasserstein regret $O_\alpha(\varepsilon_W^{\alpha/(1+\alpha)})$. This last assertion uses H\"older regularity of the \emph{future} law; experiment $F$ permits arbitrary, even atomic, historical laws. Appendix~\ref{app:ur-proof} verifies the interpolation, nonidentical-sample implementation, and all radius cases.

\section*{Conclusion}

Pricing from heterogeneous historical valuations requires balancing the performance of an established empirical rule against protection from broader future shifts. For normalized valuations, Kolmogorov heterogeneity measures discrepancies in purchase probabilities at posted-price thresholds, while one-Wasserstein heterogeneity accounts for the cost of moving valuation mass. These descriptions lead to nested ambiguity models. A Kolmogorov-regret budget preserves a benchmark guarantee as the seller seeks the best attainable Wasserstein robustness with a common pricing policy.

The H\"older safety--performance frontier quantifies the role of that budget. Tight Kolmogorov allowances yield a regularity-controlled Wasserstein rate; intermediate allowances produce inverse-budget improvement; and larger allowances reach the Wasserstein square-root rate. A related exact-value result identifies the Kolmogorov safety scale needed by policy sequences that attain the unrestricted Wasserstein-only minimax value. These findings show how the choice of a baseline guarantee affects the benefits of price adjustments under valuation shifts.

The bounded-density result establishes a complementary phenomenon. On every fixed class above the certified density threshold, maintaining a Kolmogorov budget proportional to the heterogeneity radius produces a strictly positive gap relative to the Wasserstein-only minimax value at the square-root scale. The proof combines a common-history lower certificate with an independent oracle upper certificate. It therefore demonstrates an unavoidable global adaptation cost even as sampling uncertainty vanishes.

The supporting results make the policy and informational mechanisms explicit. A randomized empirical-price undercutting rule gives finite-sample guarantees for independent nonidentical observations, while the exact three-law example shows how several futures compatible with the same history shape a global frontier. Fixed comparable unequal radii preserve the H\"older rate characterization. The analysis rests on observed full valuations and a specified baseline-risk commitment. Extending the theory to censored sales observations or learning the relevant heterogeneity description would require different information models.

\appendix
\section{Proof of the H\"older Frontier}\label{app:holder-proof}

This appendix proves Theorem~\ref{thm:holder-frontier} directly for the stochastic experiment in Section~\ref{sec:model}. Throughout, $0<\alpha<1$ is fixed, $\mathcal C=\mathcal H_{\alpha,5}$, and the common heterogeneity radius is $\varepsilon$. Constants implicit in the proof may depend on $\alpha$, but not on the budget, sample size, or choice of experiment. All price comparisons use the weak-purchase convention $V\ge p$.

\subsection{Independent, nonidentical observations and the H\"older modulus}\label{app:holder-empirical}

For independent $X_i\sim\nu_i$ define the (possibly non-H\"older) historical barycenter $\bar\nu_n=n^{-1}\sum_{i=1}^n\nu_i$, empirical measure $\widehat\nu_n=n^{-1}\sum_{i=1}^n\delta_{X_i}$, and
\[
 \eta_n=\sup_{p\in[0,1]}|r_{\widehat\nu_n}(p)-r_{\bar\nu_n}(p)|,
 \qquad \xi_n=\sup_{x\in[0,1]}|F_{\widehat\nu_n}(x)-F_{\bar\nu_n}(x)|.
\]

\begin{lemma}[Uniform non-IID empirical control]\label{lem:holder-noniid}
For every independent array of Borel laws $\nu_1,\ldots,\nu_n$ on $[0,1]$,
\begin{equation}\label{eq:holder-empirical-bound}
 \mathbb E\eta_n\le\frac4{\sqrt n},\qquad
 \mathbb E\xi_n\le\frac4{\sqrt n}.
\end{equation}
Both inequalities hold uniformly over all arrays, including atomic and nonidentically distributed ones. The empirical-revenue maximizer $\widehat p_n$ can be chosen measurably and obeys, sample by sample,
\begin{equation}\label{eq:holder-empirical-max}
 r_{\bar\nu_n}(\widehat p_n)\ge r_{\bar\nu_n}^\star-2\eta_n.
\end{equation}
\end{lemma}
\begin{proof}
For either $f_p(v)=p\mathbf 1\{v\ge p\}$ or $f_x(v)=\mathbf 1\{v\le x\}$, independence permits coordinatewise ghost-sample symmetrization, without identical distributions:
\[
 \mathbb E\sup_f\left|\frac1n\sum_{i=1}^n\bigl(f(X_i)-\mathbb Ef(X_i)\bigr)\right|
 \le\frac2n\mathbb E_{X,\sigma}\sup_f\left|\sum_{i=1}^n\sigma_i f(X_i)\right|,
\]
where the $\sigma_i$ are independent Rademacher signs. Conditional on the sample, order the valuations. Each threshold set is a prefix or suffix of that ordering (ties can only remove possible cuts). For the revenue functions, multiplication by $p\in[0,1]$ cannot increase the absolute signed sum. Consequently, each conditional supremum is bounded by the maximum absolute partial sum of the reordered signs. Doob's $L^2$ maximal inequality and Cauchy--Schwarz give
\[
 \mathbb E_\sigma\max_{0\le k\le n}\left|\sum_{i=1}^k\sigma_i\right|
 \le\bigl(4\mathbb E_\sigma|\textstyle\sum_{i=1}^n\sigma_i|^2\bigr)^{1/2}=2\sqrt n.
\]
This proves both bounds in \eqref{eq:holder-empirical-bound}. Empirical revenue attains its maximum at an observed valuation (or at zero); choosing the smallest maximizing value among these finitely many candidates is measurable. By comparing the population revenue of the selected price with its empirical revenue, and then the empirical maximum with the population optimum, one loses at most $\eta_n$ in each comparison, proving \eqref{eq:holder-empirical-max}. The survival function at $p$ is $1-F(p-)$; its empirical deviation is likewise bounded by $\xi_n$, since left limits of CDFs inherit their sup-norm bound. These arguments include prices equal to sample observations.
\end{proof}

\begin{lemma}[H\"older interpolation and barycenter control]\label{lem:holder-interpolation}
If $\mu\in\mathcal H_{\alpha,5}$ and $\nu$ is any law on $[0,1]$, then
\begin{equation}\label{eq:holder-appendix-interpolation}
 |F_\mu(x)-F_\mu(y)|\le5|x-y|^\alpha,\qquad
 d_K(\mu,\nu)\le C_\alpha^{\mathrm{int}}d_W(\mu,\nu)^{\alpha/(1+\alpha)},
 \quad C_\alpha^{\mathrm{int}}=(2\,10^{1/\alpha})^{\alpha/(1+\alpha)}.
\end{equation}
If $d(\nu_i,\mu)\le\varepsilon$ for every $i$, then $d(\bar\nu_n,\mu)\le\varepsilon$, separately for $d=K,W$, whether or not $\bar\nu_n\in\mathcal C$.
\end{lemma}
\begin{proof}
The interval-mass condition gives the first inequality and implies that $\mu$ has no atoms. For any $x$ where $F_\nu(x)-F_\mu(x)=d>0$, the inequality $F_\nu(x)\le1$ and the interval-mass bound give $d\le\mu((x,1])\le5(1-x)^\alpha$. Set $u=(d/10)^{1/\alpha}\le1-x$. For $y\in[x,x+u]$, monotonicity of $F_\nu$ and regularity of $F_\mu$ imply
\[
 F_\nu(y)-F_\mu(y)\ge d-5u^\alpha=d/2.
\]
If instead $F_\mu(x)-F_\nu(x)=d>0$, then $d\le F_\mu(x)\le5x^\alpha$, and on $[x-u,x]$ the reverse CDF difference is at least $d/2$. Thus in either case, by the integrated-CDF formula,
\[
 d_W(\mu,\nu)=\int_0^1|F_\mu(y)-F_\nu(y)|\,dy
 \ge\frac d2\left(\frac d{10}\right)^{1/\alpha}.
\]
Taking the supremum over $x$ (or a maximizing sequence) proves the stated estimate. For barycenters, $F_{\bar\nu_n}=n^{-1}\sum_iF_{\nu_i}$ gives convexity of $d_K$; the corresponding convexity of $d_W$ follows either from its integrated-CDF expression or by averaging couplings. Hence
\[
 d(\bar\nu_n,\mu)\le n^{-1}\sum_i d(\nu_i,\mu)\le\varepsilon.
\]
No assumption on the regularity of the historical marginals was used.
\end{proof}

\subsection{K-safe SAA and downward-price policies}\label{app:holder-upper}

For any two laws $\rho,\tau$ on $[0,1]$, the revenue and optimal-revenue comparisons under Kolmogorov distance are
\begin{equation}\label{eq:holder-k-revenue}
 \sup_{p\in[0,1]}|r_\rho(p)-r_\tau(p)|\le d_K(\rho,\tau),\qquad
 |r_\rho^\star-r_\tau^\star|\le d_K(\rho,\tau).
\end{equation}
Indeed, $r_\rho(p)=p(1-F_\rho(p-))$, the sup-norm CDF bound applies also to left limits, and maximizing a uniformly perturbed function changes its maximum by no more than the uniform perturbation. All population revenue maxima exist: the purchase-at-equality convention makes $p\mapsto r_\rho(p)$ upper semicontinuous on compact $[0,1]$.

\begin{lemma}[SAA at the closed safety boundary]\label{lem:holder-saa}
The policy sequence $P_n=\widehat p_n$ satisfies, for $E=F,A$,
\begin{equation}\label{eq:holder-saa-risks}
 R_{K,n}^E\le2\varepsilon+\frac8{\sqrt n},\qquad
 R_{W,n}^E\le2C_\alpha^{\mathrm{int}}\varepsilon^{\alpha/(1+\alpha)}+\frac8{\sqrt n}.
\end{equation}
Thus $R_K^E\le2\varepsilon$ and $R_W^E\le2C_\alpha^{\mathrm{int}}\varepsilon^{\alpha/(1+\alpha)}$.
\end{lemma}
\begin{proof}
For any future $\mu$ and any realization, compare $r_\mu^\star$ to $r_{\bar\nu_n}^\star$, invoke \eqref{eq:holder-empirical-max}, and compare $r_{\bar\nu_n}(\widehat p_n)$ back to $r_\mu(\widehat p_n)$. The resulting regret is bounded by $2d_K(\mu,\bar\nu_n)+2\eta_n$. In the K experiment, Lemma~\ref{lem:holder-interpolation} bounds the distance by $\varepsilon$. In the W experiment it bounds it by $C_\alpha^{\mathrm{int}}\varepsilon^{\alpha/(1+\alpha)}$, because $\mu$, but not necessarily $\bar\nu_n$, belongs to the fixed class. Take expectations, the environmental supremum, and use Lemma~\ref{lem:holder-noniid}. Then take $n\to\infty$ with $\varepsilon$ fixed. This gives feasibility at $b=2\varepsilon$ with no positive safety slack.
\end{proof}

The next comparison uses the direction of the posted-price decision. For $q\in[0,1]$ and $t>0$ let $q_t=(q-t)_+$. For any $\rho$, lowering $q$ by at most $t$ loses at most $t$ in revenue:
\begin{equation}\label{eq:holder-deflation-nominal}
 r_\rho(q_t)\ge r_\rho(q)-t.
\end{equation}
For any coupling $(X,Y)$ of $\tau$ and $\rho$, respectively, purchases with $X\ge q$ can be lost at the lower price $q_t$ only if $|X-Y|\ge t$ (when $q\ge t$). The moved mass has probability at most $\mathbb E|X-Y|/t$. When $q<t$, the right side of the inequality below is nonpositive and the assertion is immediate. Consequently,
\begin{equation}\label{eq:holder-deflation-transport}
 r_\rho(q_t)\ge r_\tau(q)-t-\frac{d_W(\rho,\tau)}t.
\end{equation}
Applying this estimate in either direction at a revenue-maximizing price and optimizing $t=\sqrt{d_W(\rho,\tau)}$ (the zero-distance case follows separately) gives
\begin{equation}\label{eq:holder-opt-w}
 |r_\rho^\star-r_\tau^\star|\le2\sqrt{d_W(\rho,\tau)}.
\end{equation}
These bounds allow arbitrary atoms and include the projection at zero.

\begin{lemma}[Statistical deterministic and randomized undercutting]\label{lem:holder-undercut}
For any fixed $t>0$, define the deterministic policy $P_n^D=(\widehat p_n-t)_+$ and the randomized policy $P_n^U=(\widehat p_n-U)_+$ with $U\sim\mathrm{Unif}[0,t]$ independent of the observations. For either $E=F,A$ and every $n$,
\begin{align}
 R_{K,n}^E(P^D)&\le2\varepsilon+t+8/\sqrt n,&
 R_{W,n}^E(P^D)&\le2\sqrt\varepsilon+t+\varepsilon/t+8/\sqrt n,
 \label{eq:holder-det-upper}\\
 R_{K,n}^E(P^U)&\le2\varepsilon+t/2+8/\sqrt n,&
 R_{W,n}^E(P^U)&\le2\sqrt\varepsilon+t/2+\varepsilon/t+8/\sqrt n.
 \label{eq:holder-rand-upper}
\end{align}
The same inequalities, without $8/\sqrt n$, hold for the asymptotic risks.
\end{lemma}
\begin{proof}
For the deterministic policy, \eqref{eq:holder-deflation-nominal}, \eqref{eq:holder-empirical-max}, and \eqref{eq:holder-k-revenue} yield, conditional on the data,
\[
 \ell_\mu(\delta_{P_n^D})\le2\varepsilon+t+2\eta_n
\]
in a K environment. In a W environment, \eqref{eq:holder-opt-w} and \eqref{eq:holder-deflation-transport} instead give
\[
 \ell_\mu(\delta_{P_n^D})\le2\sqrt\varepsilon+t+\varepsilon/t+2\eta_n.
\]
For the uniform policy, lowering $q$ by $U$ costs at most $\min\{U,q\}\le U$ under any fixed law. Hence
\[
 \mathbb E_U r_\rho((q-U)_+)\ge r_\rho(q)-t/2.
\]
For a fixed $q$, its smoothed revenue as a function of valuation is
\[
 g_{q,t}(v)=\mathbb E_U\bigl[(q-U)_+\mathbf1\{v\ge(q-U)_+\}\bigr]
 =\frac1t\int_{(q-t)_+}^{\min\{q,v\}}z\,dz
\]
when $v\ge(q-t)_+$, and zero otherwise. This function is continuous, piecewise differentiable with slopes bounded by $1/t$ (because $0\le z\le1$); the formula also covers $q<t$, when $U\ge q$ produces a zero price. By any coupling, or Wasserstein duality,
\begin{equation}\label{eq:holder-smoothed-revenue}
 \left|\mathbb E_Ur_\rho((q-U)_+)-\mathbb E_Ur_\tau((q-U)_+)\right|
 \le\frac{d_W(\rho,\tau)}t.
\end{equation}
Apply the $t/2$ nominal estimate and \eqref{eq:holder-smoothed-revenue} with $q=\widehat p_n$, conditionally on the sample, followed by \eqref{eq:holder-empirical-max} and \eqref{eq:holder-opt-w}, to prove the W bound in \eqref{eq:holder-rand-upper}. For its K bound, use \eqref{eq:holder-k-revenue} on both sides of the nominal comparison. Finally take the sample expectation, then the supremum over admissible independent historical arrays and futures, and use $2\sup\mathbb E\eta_n\le8/\sqrt n$. Every width is fixed at the relevant $\varepsilon,b$ before $n\to\infty$, and neither policy uses the identity of the metric world.
\end{proof}

\subsection{The partial-mass witness: geometry and every-price inequality}\label{app:holder-lower-geometry}

Set
\begin{equation}\label{eq:holder-witness-normalization}
 a=\frac12,\qquad
 h=(4^{-\alpha}\varepsilon)^{1/(1+\alpha)},\qquad
 \delta=(4h)^\alpha,\qquad \delta h=\varepsilon.
\end{equation}
As $\varepsilon\downarrow0$ with fixed $\alpha<1$, we have $h,\delta\to0$, $\delta/h\to\infty$, $h^2/\varepsilon=h/\delta\to0$, and $\delta/\sqrt\varepsilon\to\infty$. Define
\begin{equation}\label{eq:holder-witness-distributions}
 \nu=\delta\,\mathrm{Unif}[a,a+h]+(1-\delta)\,\mathrm{Unif}[a+h,1].
\end{equation}
For $0<m\le\delta/4$ put
\begin{equation}\label{eq:holder-witness-parameters}
 s=\frac{mh}{\delta},\qquad D=\frac\varepsilon m,\qquad
 Q=am-D-s,\qquad t_0=D-s.
\end{equation}
Let $\mu_m$ be obtained from $\nu$ by translating left by $D$ precisely the mass $m$ that is uniform on $[a,a+s]$, leaving the other mass unchanged. This is well-defined as a probability distribution because $\nu([a,a+s])=(\delta/h)s=m$. We require $D>s$ and $D\le a$, in addition to $0<m\le\delta/4$: the shifted interval $[a-D,a-D+s]$ then lies inside $[0,a)$ and is disjoint from the remaining rare interval $[a+s,a+h]$. Every parameter choice used below meets these restrictions for sufficiently small $\varepsilon$.

\begin{lemma}[Admissibility and exact transport]\label{lem:holder-witness-valid}
For each fixed $0<\alpha<1$ and all sufficiently small $\varepsilon$, uniformly for $m$ satisfying the stated geometric restrictions,
\begin{equation}\label{eq:holder-witness-valid}
 \nu,\mu_m\in\mathcal H_{\alpha,5},\qquad
 d_W(\nu,\mu_m)=mD=\varepsilon,\qquad
 r_\nu^\star=a,\quad r_{\mu_m}^\star\ge a-D.
\end{equation}
\end{lemma}
\begin{proof}
The rare-component density is $\rho=\delta/h=4^\alpha h^{\alpha-1}$, and the background density is $q_0=(1-\delta)/(1/2-h)$. For fixed $\alpha<1$ and sufficiently small $\varepsilon$ we can simultaneously require
\begin{equation}\label{eq:holder-geometry-restrictions}
 h<1/8,\quad0<\delta<1/4,\quad q_0>0,\quad\rho>2,\quad
 q_0\le3,\quad \rho\ge q_0,\quad
 4^\alpha+3h^{1-\alpha}<5.
\end{equation}
Both laws have total rare mass $\delta$ spread at density $\rho$ on one or two intervals, and the remaining mass has density $q_0$ on $[a+h,1]$. Thus for \emph{every} interval $I\subseteq[0,1]$ of length $L$,
\[
 \nu(I),\mu_m(I)\le\min\{\rho L,\delta+q_0L\}.
\]
For $0<L\le h$, the first bound is $\rho L=4^\alpha h^{\alpha-1}L\le4^\alpha L^\alpha<5L^\alpha$. For $h\le L\le1$, the ratio of the second bound to $L^\alpha$ is
\[
 f(L)=\delta L^{-\alpha}+q_0L^{1-\alpha}.
\]
Its derivative has the sign of $(1-\alpha)q_0L-\alpha\delta$, so its maximum on $[h,1]$ is attained at an endpoint. The endpoint values satisfy
\[
 f(h)=4^\alpha+q_0h^{1-\alpha}\le4^\alpha+3h^{1-\alpha}<5,
 \qquad f(1)=\delta+q_0<5.
\]
Length-zero intervals have mass zero. Hence both measures satisfy the class condition for all intervals, including those meeting the separated portions of the rare mass. The bounds are independent of $m$.

Couple the unchanged mass identically and map each point of $[a,a+s]$ to its translate by $-D$. This coupling has cost $mD$. Conversely, the means differ by exactly $mD$, and the test function $v\mapsto v$ is 1-Lipschitz; hence every coupling has expected absolute displacement at least $mD$. The transport cost is therefore \emph{exactly} $mD=\varepsilon$.

Every valuation under $\nu$ is at least $a$. Below $a$, its revenue is $p$; on $[a,a+h]$ the revenue is $p[1-\rho(p-a)]$, whose derivative is at most $1-\rho a<0$ when $\rho>2$; above $a+h$ it is $q_0p(1-p)$, nonincreasing because $a+h>1/2$. Thus $r_\nu^\star=r_\nu(a)=a$. Every valuation under $\mu_m$ is at least $a-D\ge0$, so pricing at $a-D$ sells to all customers and yields $a-D$. This proves the final claims without assuming that $a-D$ is exactly optimal for $\mu_m$.
\end{proof}

\begin{lemma}[Every-price common-history certificate]\label{lem:holder-all-prices}
Under the restrictions of Lemma~\ref{lem:holder-witness-valid}, suppose also $Q>0$ and define $\lambda=\max\{1,Q/t_0\}$, where $t_0=D-s>0$. Then for every $p\in[0,1]$,
\begin{equation}\label{eq:holder-appendix-all-prices}
 \ell_{\mu_m}(\delta_p)+\lambda\ell_\nu(\delta_p)\ge Q.
\end{equation}
Consequently, for \emph{every} sequence of measurable, potentially labeled and randomized pricing policies, both $E=F,A$ satisfy
\begin{equation}\label{eq:holder-statistical-lower}
 R_W^E(\pi)+\lambda R_K^E(\pi)\ge Q,
 \qquad R_W^E(\pi)\ge\frac Q{1+\lambda}.
\end{equation}
\end{lemma}
\begin{proof}
For $p\le a-D+s<a$, the original future has regret $\ell_\nu(\delta_p)=a-p\ge D-s=t_0$; by definition $\lambda t_0\ge Q$, and the other regret is nonnegative. For $a-D+s<p<a+s$, all translated mass lies below the price and all the unshifted mass lies at or above it, so $r_{\mu_m}(p)=p(1-m)\le(a+s)(1-m)$. For $p\ge a+s$, the shifted mass and its original location are both below the price, whence $r_{\mu_m}(p)=r_\nu(p)$. Since $r_\nu$ is nonincreasing for $p\ge a$, this is at most $r_\nu(a+s)=(a+s)(1-m)$. Thus on the remaining prices,
\[
 \ell_{\mu_m}(\delta_p)\ge a-D-(a+s)(1-m)
 =am-D-s+ms\ge Q.
\]
This covers $p=a-D+s$ by the first case and $p=a+s$ by the last. At endpoints of the translated intervals there are no atoms; the weak-purchase convention changes none of these computations. Averaging the inequality over any probability distribution of prices preserves it.

For fixed $n$, construct a K-world with future $\nu$ and historical marginals $\nu_i=\nu$ and a W-world with future $\mu_m$ and the \emph{same} marginals $\nu_i=\nu$. The K-distance in the first world is zero, and the W-distance in the second is exactly $\varepsilon$. By Lemma~\ref{lem:holder-witness-valid}, \emph{both} future and historical laws belong to $\mathcal C$, so the worlds are permissible separately in experiments $A$ and $F$. Their ordered sample laws coincide and equal $\nu^{\otimes n}$, regardless of whether a policy treats sample indices differently. Integrating \eqref{eq:holder-appendix-all-prices} over this shared sample law and its output kernel gives
\[
 R_{W,n}^E(\pi)+\lambda R_{K,n}^E(\pi)\ge Q.
\]
Furthermore, the K-world with future $\nu$ is itself W-admissible (distance zero); both risks from the two particular worlds are consequently bounded above by $R_{W,n}^E(\pi)$. Hence $(1+\lambda)R_{W,n}^E(\pi)\ge Q$. Taking $\limsup_{n\to\infty}$ in the latter inequality gives the second assertion of \eqref{eq:holder-statistical-lower}. For the first, $R_{W,n}^E\ge Q-\lambda R_{K,n}^E$ yields
\[
 R_W^E=\limsup_nR_{W,n}^E
 \ge Q-\lambda\liminf_nR_{K,n}^E
 \ge Q-\lambda R_K^E.
\]
In particular no uniform convergence, IID restriction on the policy, or population-game interchange is needed for the lower bounds.
\end{proof}

\subsection{Three valid witness choices and uniform stitching}\label{app:holder-stitching}

We may choose a single radius threshold $\varepsilon_0(\alpha)>0$ so that all inequalities in \eqref{eq:holder-geometry-restrictions} and the following conditions hold:
\begin{equation}\label{eq:holder-uniform-smallness}
 \frac\delta h\ge68,\qquad
 \frac{h^2}{\varepsilon}\le\frac1{128},\qquad
 \frac\delta{\sqrt\varepsilon}\ge32,\qquad
 h\le\frac{\sqrt\varepsilon}{16},\qquad
 4\varepsilon\le h,
 \qquad 2\varepsilon+\sqrt{\varepsilon/2}\le\sqrt\varepsilon.
\end{equation}
Such a threshold exists because $0<\alpha<1$ is fixed. The lower witnesses use only their individually valid ranges; no choice of $m$ is extrapolated to other budgets.

\paragraph{Low budgets: $2\varepsilon\le b\le h$.}
Choose $m=\delta/4$, so $s=h/4$, $D=4h$, and $t_0=15h/4$. Then $m\le\delta/4$, $D>s$, and $D\le a$ because $h\le1/8$. The lower certificate has
\[
 Q=\frac\delta8-\frac{17h}4\ge\frac\delta{16}>0,
 \qquad Q\ge t_0,
 \qquad\lambda=\frac Q{t_0},
\]
where $\delta/h\ge68$ suffices. By Lemma~\ref{lem:holder-all-prices}, any policy with $R_K^E\le b$ satisfies
\begin{equation}\label{eq:holder-low-lower}
 R_W^E\ge Q-\lambda b
 \ge Q\left(1-\frac4{15}\right)
 \ge\frac{11}{240}\delta.
\end{equation}

\paragraph{Intermediate budgets: $h\le b\le\sqrt\varepsilon/16$.}
Choose $m=\varepsilon/(4b)$, hence $D=4b$ and $s=h^2/(4b)$. Since $b\ge h$, $m\le\delta/4$ and $s\le h/4$; since $b\le\sqrt\varepsilon/16$, $D\le\sqrt\varepsilon/4\le a$ for small radius. Also $t_0=4b-h^2/(4b)\ge15b/4>0$. The gap satisfies
\[
 Q=\frac\varepsilon{8b}-4b-\frac{h^2}{4b}
 =\frac\varepsilon{8b}\left(1-\frac{32b^2}{\varepsilon}-\frac{2h^2}{\varepsilon}\right)
 \ge\frac{3\varepsilon}{32b}.
\]
Here $32b^2/\varepsilon\le1/8$ and $2h^2/\varepsilon\le1/64$. Furthermore, $Q\ge24b\ge t_0$ in this range, so $\lambda=Q/t_0$ and $\lambda b\le4Q/15$. Thus
\begin{equation}\label{eq:holder-mid-lower}
 R_W^E\ge\frac{11}{15}Q\ge\frac{11}{160}\frac\varepsilon b.
\end{equation}
The inverse-budget bound is obtained from an actual transported-mass witness, not by interpolating endpoint inequalities.

\paragraph{High budgets and the unconditional W floor.}
Choose $m=8\sqrt\varepsilon$, giving
\[
 D=\frac{\sqrt\varepsilon}{8},\qquad
 s=\frac{8h^2}{\sqrt\varepsilon}\le\frac{\sqrt\varepsilon}{16}.
\]
The condition $\delta/\sqrt\varepsilon\ge32$ ensures $m\le\delta/4$; $D>s$, $D\le a$, and the shifted support is nonnegative. Directly,
\[
 Q=4\sqrt\varepsilon-D-s\ge\frac{61}{16}\sqrt\varepsilon>0,
 \qquad t_0\ge\frac{\sqrt\varepsilon}{16},
 \qquad\lambda=\max\{1,Q/t_0\}\le64.
\]
The second inequality in \eqref{eq:holder-statistical-lower} now yields for \emph{every} policy, without a K-budget assumption,
\begin{equation}\label{eq:holder-high-lower}
 R_W^E\ge\frac{61}{1040}\sqrt\varepsilon.
\end{equation}
This lower bound uses the same two W-admissible futures and their indistinguishable history; it is not an assertion about an exact Wasserstein minimax coefficient.

\paragraph{Upper policies and the closed budget boundary.}
Write $\beta=\alpha/(1+\alpha)$ and $k_\alpha=4^{\alpha/(1+\alpha)}$, so
\[
 \delta=k_\alpha\varepsilon^\beta,\qquad
 h=k_\alpha^{-1}\varepsilon^{1/(1+\alpha)},\qquad \varepsilon/h=\delta.
\]
For $2\varepsilon\le b\le h$ use SAA. Lemma~\ref{lem:holder-saa} proves $R_K^E\le2\varepsilon\le b$ and $R_W^E\le2C_\alpha^{\mathrm{int}}\varepsilon^\beta$. For $h\le b\le\sqrt\varepsilon$, take the deterministic price reduction $t=b-2\varepsilon>0$. By $2\varepsilon/h\to0$, our radius threshold ensures $t\ge b/2$, hence
\[
 R_K^E(P^D)\le b,\qquad
 R_W^E(P^D)\le2\sqrt\varepsilon+b+\frac{2\varepsilon}{b}
 \le5\frac\varepsilon b.
\]
For $b\ge\sqrt\varepsilon$, use randomized undercutting with $t=\sqrt{2\varepsilon}$. By \eqref{eq:holder-uniform-smallness},
\[
 R_K^E(P^U)\le2\varepsilon+\sqrt{\varepsilon/2}\le\sqrt\varepsilon\le b,
 \qquad R_W^E(P^U)\le(2+\sqrt2)\sqrt\varepsilon.
\]
All are explicitly sample-dependent policies, and all bounds are obtained by taking $n\to\infty$ before changing $\varepsilon$.

\paragraph{Uniform comparison for every $b\ge2\varepsilon$.}
Put $\Phi=\max\{\sqrt\varepsilon,\min\{\varepsilon^\beta,\varepsilon/b\}\}$. In the low range, $\varepsilon/b\ge\varepsilon/h=\delta\ge\varepsilon^\beta$ and $\varepsilon^\beta\ge\sqrt\varepsilon$ for sufficiently small $\varepsilon$, so $\Phi=\varepsilon^\beta$; the SAA upper bound and \eqref{eq:holder-low-lower} match it up to constants depending only on $\alpha$. For $h\le b\le\sqrt\varepsilon/16$, we have $\Phi\le\varepsilon/b$, so \eqref{eq:holder-mid-lower} bounds it below. For $b\ge\sqrt\varepsilon/16$, $\Phi\le16\sqrt\varepsilon$, so \eqref{eq:holder-high-lower} bounds it below. These intervals cover all eligible budgets, including their endpoints.

For the upper comparison in $h\le b\le\sqrt\varepsilon$, note that $\sqrt\varepsilon\le\varepsilon/b\le\delta=k_\alpha\varepsilon^\beta$; therefore $\varepsilon/b\le k_\alpha\Phi$. At $b\ge\sqrt\varepsilon$, $\Phi=\sqrt\varepsilon$. Thus a single positive lower constant and a single finite upper constant depending only on the fixed $\alpha$ bound $G_E(b)$ on all three ranges. The same policies work in experiment $F$ because only the future needs to be H\"older, whereas the constructed history and both futures in each lower witness belong to $\mathcal H_{\alpha,5}$ and hence are admissible even in $A$. This proves \eqref{eq:holder-frontier} uniformly for $E\in\{F,A\}$ and $b\ge2\varepsilon$.

Finally, the logical limit order has been preserved throughout: finite-$n$ bounds first take the expectation over samples and randomized prices, then the supremum over environments; their uniform $O(n^{-1/2})$ terms vanish in the fixed-radius $\limsup_{n\to\infty}$. For each fixed $\varepsilon,b$ the displayed policies are feasible sequences; taking the infimum over all eligible sequences cannot weaken the upper certificate or evade the arbitrary-policy lower certificates. Only then is $\varepsilon\downarrow0$ invoked to select a common $\varepsilon_0(\alpha)$. There is no interchange of the sample-size and radius limits, no assumption that histories are IID in the upper proof, and no reduction of the policy class in the lower proof.

\section{Proof of Exact Wasserstein-Oracle Access}\label{app:oracle-proof}

This appendix proves Theorem~\ref{thm:exact-oracle-access} in the policy-sequence experiment of Section~\ref{sec:model}. We first establish the universal Wasserstein benchmark for independent, potentially nonidentically distributed full valuations, then justify exact minimax and minimum-budget attainment. The lower separation reuses the all-price certificate already proved in Appendix~\ref{app:holder-proof}; its interval-mass and transport arguments are not repeated.

\subsection{Universal Wasserstein benchmark}\label{app:oracle-universal}

\begin{lemma}[Smoothed-price Wasserstein bound]\label{lem:oracle-universal}
For any nonempty $\mathcal C\subseteq\mathcal P$, either experiment $E\in\{F,A\}$, and $0<\varepsilon\le1/2$, an admissible sample-dependent randomized policy sequence satisfies
\begin{equation}\label{eq:oracle-smoothing-finite}
 R_{W,n}^E(\pi;\mathcal C,\varepsilon)
 \le\sqrt{2\varepsilon}
 +\frac{1}{2\sqrt{2\varepsilon}\sqrt n},\qquad n\ge1.
\end{equation}
Consequently, $W_E^\star(\mathcal C,\varepsilon)\le\sqrt{2\varepsilon}$.
\end{lemma}
\begin{proof}
Fix a bandwidth $t\in(0,1]$. For a probability law $\rho$ and an endpoint $z\in[t,1]$, put
\[
 S_t(\rho,z)=\frac1t\int_{z-t}^{z}r_\rho(p)\,dp.
\]
Given the full sample $X_{1:n}$, let $\widehat\nu_n=n^{-1}\sum_{i=1}^n\delta_{X_i}$, choose a maximizer
$\widehat z_n\in\arg\max_{z\in[t,1]}S_t(\widehat\nu_n,z)$,
and post a price uniformly on $[\widehat z_n-t,\widehat z_n]$. The map $z\mapsto S_t(\widehat\nu_n,z)$ is continuous on the compact interval $[t,1]$, since it is the moving integral of a bounded measurable revenue function. Joint measurability in the sample and $z$, followed by choosing the smallest maximizer, gives a measurable stochastic kernel. No distributional information is used by this policy beyond $\varepsilon$ and the observed sample.

Lowering a price cannot decrease the probability of sale, so for $0\le p\le z$,
$r_\mu(p)\ge r_\mu(z)-(z-p)$. Averaging over $p\in[z-t,z]$ gives
\begin{equation}\label{eq:oracle-smoothing-bias}
 S_t(\mu,z)\ge r_\mu(z)-t/2.
\end{equation}
For $v\in[0,1]$ define
\[
 g_{z,t}(v)=\frac1t\int_{z-t}^{z}p\,\mathbf1\{v\ge p\}\,dp,
 \qquad S_t(\rho,z)=\int g_{z,t}(v)\,\rho(dv).
\]
For two endpoints $z,z'\in[t,1]$, the difference $g_{z,t}-g_{z',t}$ is absolutely continuous in $v$ and has derivative, almost everywhere,
\[
 \frac vt\bigl(\mathbf1_{[z-t,z]}(v)-\mathbf1_{[z'-t,z']}(v)\bigr).
\]
Its absolute value is at most $1/t$; importantly, the difference of the indicators has magnitude at most one. Coupling two valuation laws therefore yields the \emph{pairwise} comparison
\begin{equation}\label{eq:oracle-smoothing-comparison}
 \left|\bigl[S_t(\mu,z)-S_t(\mu,z')\bigr]
       -\bigl[S_t(\rho,z)-S_t(\rho,z')\bigr]\right|
 \le\frac{d_W(\mu,\rho)}t.
\end{equation}
This comparison avoids paying two separate Wasserstein errors for the candidate and selected prices.

Choose a future-revenue maximizer $p_\mu^\star$ as in \eqref{eq:model-revenue}. If $p_\mu^\star\ge t$, empirical optimality of $\widehat z_n$ and \eqref{eq:oracle-smoothing-bias}--\eqref{eq:oracle-smoothing-comparison}, with $\rho=\widehat\nu_n$, give, conditional on the observed sample,
\begin{equation}\label{eq:oracle-smoothing-regret}
 \begin{split}
 \ell_\mu\bigl(\mathrm{Unif}[\widehat z_n-t,\widehat z_n]\bigr)
 &=r_\mu^\star-S_t(\mu,\widehat z_n)\\
 &\le t/2+S_t(\mu,p_\mu^\star)-S_t(\mu,\widehat z_n)\\
 &\le t/2+d_W(\mu,\widehat\nu_n)/t.
 \end{split}
\end{equation}
If $p_\mu^\star<t$, then $r_\mu^\star\le p_\mu^\star<t$, and nonnegative pricing revenue bounds the regret by $t$. These two cases use a fixed property of the future law, not a data-dependent event.

In any Wasserstein-admissible environment, set $\bar\nu_n=n^{-1}\sum_{i=1}^n\nu_i$. The componentwise heterogeneity restrictions imply
\[
 d_W(\mu,\bar\nu_n)
 \le\frac1n\sum_{i=1}^nd_W(\mu,\nu_i)\le\varepsilon.
\]
Independence, without an identical-distribution assumption, implies at each $x\in[0,1]$ that
\[
 \mathbb E\bigl|F_{\widehat\nu_n}(x)-F_{\bar\nu_n}(x)\bigr|
 \le\sqrt{\operatorname{Var}\!\bigl(F_{\widehat\nu_n}(x)\bigr)}
 \le\frac1{2\sqrt n}.
\]
Integrating over $x$ in the one-dimensional Wasserstein formula gives
\begin{equation}\label{eq:oracle-smoothing-empirical}
 \mathbb E\,d_W(\widehat\nu_n,\bar\nu_n)\le\frac1{2\sqrt n},
 \qquad \mathbb E\,d_W(\mu,\widehat\nu_n)
 \le\varepsilon+\frac1{2\sqrt n}.
\end{equation}
Take $t=\sqrt{2\varepsilon}\le1$. In the case $p_\mu^\star\ge t$, the expected bound in \eqref{eq:oracle-smoothing-regret} is at most
$t/2+\varepsilon/t+1/(2t\sqrt n)=t+1/(2t\sqrt n)$; in the other case it is at most $t$. These estimates hold uniformly over all admissible futures and independent component arrays, for both experiments. Taking the environmental supremum after expectation proves \eqref{eq:oracle-smoothing-finite}. At each fixed $\varepsilon>0$, take $\limsup_{n\to\infty}$ and then the infimum over sequences to obtain $W_E^\star\le\sqrt{2\varepsilon}$.
\end{proof}

\subsection{Attainment in the asymptotic policy-sequence class}\label{app:oracle-attainment}

\begin{lemma}[Exact minimax and least-safety attainment]\label{lem:oracle-attainment}
For any nonempty future class $\mathcal C$, fixed radius $\varepsilon\ge0$, and fixed experiment $E\in\{F,A\}$, there is an admissible sequence $\pi^W$ with
$R_W^E(\pi^W;\mathcal C,\varepsilon)=W_E^\star(\mathcal C,\varepsilon)$.
Moreover, there is an admissible sequence $\pi^{W,K}$ that attains both
\begin{equation}\label{eq:oracle-attained-pair}
 R_W^E(\pi^{W,K};\mathcal C,\varepsilon)=W_E^\star(\mathcal C,\varepsilon),
 \qquad
 R_K^E(\pi^{W,K};\mathcal C,\varepsilon)=b_E^W(\mathcal C,\varepsilon).
\end{equation}
In particular, the minimum in \eqref{eq:model-access-budget} is well defined, and
$G_E(b_E^W;\mathcal C,\varepsilon)=W_E^\star(\mathcal C,\varepsilon)$.
\end{lemma}
\begin{proof}
In this proof hold $\mathcal C,\varepsilon,E$ fixed and abbreviate the finite-sample risks by $r_{d,n}(\pi_n)$ and the sequence risks by $R_d(\pi)=\limsup_nr_{d,n}(\pi_n)$. Each $r_{d,n}$ depends only on the single $n$-sample stochastic kernel $\pi_n$. The policy class permits an arbitrary measurable kernel at every $n$, with no compatibility constraint across sample sizes. All risks lie in $[0,1]$, so the infima below are finite.

\emph{First diagonalization (the Wasserstein minimax infimum).} For each integer $j\ge1$, the definition of $W_E^\star$ supplies a full sequence $\pi^{(j)}$ satisfying
\[
 R_W(\pi^{(j)})\le W_E^\star+\frac1j.
\]
By the definition of a limsup, there is an integer $N_j'$ such that
\[
 r_{W,n}(\pi_n^{(j)})\le W_E^\star+\frac2j
 \qquad\text{for every }n\ge N_j'.
\]
Choose strictly increasing integers $N_j\ge\max\{j,N_j'\}$ recursively. For $N_j\le n<N_{j+1}$ set $\pi_n^W=\pi_n^{(j)}$; for the finitely many $n<N_1$, use any measurable pricing kernel. Each stage is a legitimate, possibly $n$-dependent kernel, and the stage index $j(n)$ tends to infinity with $n$. Hence
\[
 R_W(\pi^W)\le\limsup_{n\to\infty}
 \left(W_E^\star+\frac{2}{j(n)}\right)=W_E^\star.
\]
The reverse inequality holds for every policy sequence by the definition of the infimum. Thus $R_W(\pi^W)=W_E^\star$; no single finite-$n$ minimizer was assumed.

\emph{Second diagonalization (least Kolmogorov risk at exact $W$ optimality).} The set of exact $W$-optimal sequences is now nonempty. Write
\[
 \beta=\inf\{R_K(\pi):\ R_W(\pi)=W_E^\star\}.
\]
For each $j\ge1$ choose an exact $W$-optimal sequence $\rho^{(j)}$ such that
\[
 R_W(\rho^{(j)})=W_E^\star,\qquad
 R_K(\rho^{(j)})\le\beta+\frac1j.
\]
By the two limsup statements, choose a common integer $M_j'$ after which, \emph{simultaneously for every} $n\ge M_j'$,
\[
 r_{W,n}(\rho_n^{(j)})\le W_E^\star+\frac1j,
 \qquad
 r_{K,n}(\rho_n^{(j)})\le\beta+\frac2j.
\]
Choose strictly increasing $M_j\ge\max\{j,M_j'\}$ and set $\pi_n^{W,K}=\rho_n^{(j)}$ for $M_j\le n<M_{j+1}$, with an arbitrary kernel for $n<M_1$. The same sequence of stage choices enforces both bounds, yielding
\[
 R_W(\pi^{W,K})\le W_E^\star,
 \qquad R_K(\pi^{W,K})\le\beta.
\]
The first is equality by minimax optimality; the second is equality by the definition of $\beta$ among exactly $W$-optimal sequences. Thus $\beta$ is attained and equals the minimum $b_E^W$ in \eqref{eq:model-access-budget}. The attained sequence is $K$-feasible at budget $b_E^W$ and has $W$-risk $W_E^\star$, so $G_E(b_E^W)\le W_E^\star$; the opposite inequality follows because the frontier minimizes over a subset of all policy sequences.
\end{proof}

\subsection{The uniform separation and proof of Theorem~\ref{thm:exact-oracle-access}}\label{app:oracle-separation}

\begin{proof}[Proof of Theorem~\ref{thm:exact-oracle-access}]
Fix $0<\alpha<1$, $\mathcal C=\mathcal H_{\alpha,5}$, and $E\in\{F,A\}$. Take $\varepsilon>0$ below the fixed-$\alpha$ threshold in \eqref{eq:holder-uniform-smallness} and small enough that the universal construction in Lemma~\ref{lem:oracle-universal} has $t=\sqrt{2\varepsilon}\le1$. The high-budget witness in Appendix~\ref{app:holder-stitching} selects
\[
 m=8\sqrt\varepsilon,
 \qquad D=\frac{\sqrt\varepsilon}{8},
 \qquad s=\frac{8h^2}{\sqrt\varepsilon}\le\frac{\sqrt\varepsilon}{16},
\]
with $h$ and $\delta$ as in \eqref{eq:holder-witness-normalization}. That appendix establishes all admissibility restrictions for $\nu,\mu_m\in\mathcal H_{\alpha,5}$ and exact transport distance $d_W(\nu,\mu_m)=\varepsilon$. In particular, for the parameters of Lemma~\ref{lem:holder-all-prices},
\[
 Q=\frac m2-D-s\ge\frac{61}{16}\sqrt\varepsilon,
 \quad t_0=D-s\ge\frac{\sqrt\varepsilon}{16},
 \quad \lambda=\max\{1,Q/t_0\}\le64,
\]
where the last inequality uses $Q\le m/2=4\sqrt\varepsilon$. The all-price lower inequality \eqref{eq:holder-statistical-lower}, already proved for every randomized, sample-dependent policy sequence with identical historical laws in both metric worlds, therefore gives
\[
 R_W^E(\pi)+64R_K^E(\pi)
 \ge R_W^E(\pi)+\lambda R_K^E(\pi)
 \ge Q\ge\frac{61}{16}\sqrt\varepsilon
\]
for \emph{every} $\pi\in\Pi$. The first step uses $R_K^E(\pi)\ge0$. This is \eqref{eq:exact-oracle-separation}; its quantifiers are uniform over the full policy class, not just a family of undercutting rules. Both future laws and the common historical law lie in $\mathcal H_{\alpha,5}$, so the argument applies separately to $F$ and $A$.

By Lemma~\ref{lem:oracle-attainment}, select an \emph{actually exact} Wasserstein-minimax sequence $\pi^{W,K}$ attaining the minimum $b_E^W$. Substituting $R_W^E(\pi^{W,K})=W_E^\star$ into the preceding inequality and invoking Lemma~\ref{lem:oracle-universal} yields
\[
 64b_E^W
 \ge\frac{61}{16}\sqrt\varepsilon-W_E^\star
 \ge\left(\frac{61}{16}-\sqrt2\right)\sqrt\varepsilon.
\]
The coefficient is strictly positive. For the opposite bound, the inclusion
$\mathfrak E_{K,n}^E\subseteq\mathfrak E_{W,n}^E$ in \eqref{eq:model-environments} ensures $R_K^E(\pi)\le R_W^E(\pi)$ for every sequence. In particular,
\[
 b_E^W=R_K^E(\pi^{W,K})
 \le R_W^E(\pi^{W,K})=W_E^\star
 \le\sqrt{2\varepsilon}.
\]
These are the two inequalities of \eqref{eq:exact-oracle-access-bounds}. Their constants are independent of $E$ and $\varepsilon$ once $\alpha$ is fixed, so the $\Theta_\alpha(\sqrt\varepsilon)$ statement follows. All risks were defined using the sample expectation, the environmental supremum, and then the $n\to\infty$ limsup at each fixed $\varepsilon$; only afterward is the small-radius restriction applied. None of the diagonal sequences is claimed to minimize the risk at any particular finite sample size, and the universal smoothing bound is not identified with the exact Wasserstein minimax coefficient.
\end{proof}

\section{Proof of the Bounded-Density Adaptation Gap}\label{app:density-gap}

We prove Theorem~\ref{thm:density-gap} using the risk and environment definitions in \eqref{eq:model-environments}--\eqref{eq:model-adaptation-gap}. Throughout, $B>17/8$ and $L\ge2$ are fixed, and $\varepsilon\downarrow0$ only after the complete $n\to\infty$ statistical risk has been formed. We first work under the weaker condition $B\ge2$ for the constrained lower certificate. All constructed distributions are atomless, so their revenue functions are continuous; all historical arrays used in the lower bound are admissible in both experiments.

\subsection{Saturated two-peak construction and global safety forcing}\label{app:density-lower}

\begin{lemma}[Two-peak constrained lower certificate]\label{lem:density-two-peak}
For each fixed $B\ge2$, $L\ge2$ and $E\in\{F,A\}$,
\begin{equation}\label{eq:density-constrained-lower}
 \liminf_{\varepsilon\downarrow0}
 \frac{A_L^E(\mathcal D_B,\varepsilon)}{\sqrt\varepsilon}
 \ge \sqrt B\,P_B=C_B.
\end{equation}
The bound holds for all measurable randomized pricing kernels, including those that depend on sample order and are evaluated under arbitrary independent non-IID environments.
\end{lemma}
\begin{proof}
Fix $\varepsilon>0$ sufficiently small and put $m=\varepsilon^{3/4}$. Let $a=a_\varepsilon$ be the larger root in $(1/2,1)$ of
\begin{equation}\label{eq:density-a-root}
 Ba(1-a)=\frac1{4B}+m,
 \qquad
 a=\frac{1+\sqrt{1-B^{-2}-4m/B}}2.
\end{equation}
For fixed $B\ge2$, the square root is real and positive when $m<(B^2-1)/(4B)$. As $\varepsilon\downarrow0$, $a\uparrow P_B$ from below. Define
\begin{equation}\label{eq:density-center-parameters}
 c=B(1-a),\qquad d=\frac{1-c}{B}=a+\frac1B-1,
 \qquad b=\frac1{2B},\qquad w=\sqrt{\frac\varepsilon B}.
\end{equation}
Since $B\ge2$ and $P_B>1/2$, for all sufficiently small $\varepsilon$,
\begin{equation}\label{eq:density-support-conditions}
 \frac12<a<1,\quad 0<c<\frac12,\quad
 0<b<d<a-w<a<a+w<1,\quad Bw<1.
\end{equation}
Indeed, at $m=0$ we have $B(1-P_B)=1/[2B(1+\sqrt{1-B^{-2}})]<1/2$; the other strict inequalities follow from $d=a+1/B-1<a$, $b<d$ when $c<1/2$, and $w\to0$. Hence these are genuine support and radius feasibility conditions, rather than informal small-interval approximations.

Define the historical law $H$ and alternative future law $J$ by their densities:
\begin{equation}\label{eq:density-two-peak-laws}
 \begin{split}
 f_H(x)&=B\mathbf1_{[0,d]}(x)+B\mathbf1_{[a,1]}(x),\\
 f_J(x)&=B\mathbf1_{[0,d]}(x)+B\mathbf1_{[a-w,a]}(x)
                 +B\mathbf1_{[a+w,1]}(x).
 \end{split}
\end{equation}
By \eqref{eq:density-support-conditions}, the displayed intervals are disjoint except possibly at endpoints of Lebesgue measure zero and lie inside $[0,1]$. Both densities are nonnegative and at most $B$ almost everywhere. Their masses agree and equal one:
\begin{equation}\label{eq:density-mass-check}
 \int_0^1 f_H\,dx=Bd+B(1-a)=1,
 \qquad
 \int_0^1 f_J\,dx=Bd+Bw+B(1-a-w)=1.
\end{equation}
Thus $H,J\in\mathcal D_B$. The second law moves a block of mass $Bw$ from $[a,a+w]$ to $[a-w,a]$, preserving density saturation. Direct integration gives the exact CDF difference
\begin{equation}\label{eq:density-cdf-triangle}
 F_J(x)-F_H(x)=
 \begin{cases}
 B(x-a+w),&a-w\le x\le a,\\
 B(a+w-x),&a\le x\le a+w,\\
 0,&\text{otherwise}.
 \end{cases}
\end{equation}
It is nonnegative and triangular, with base $2w$ and height $Bw$, so the one-dimensional transport identity in \eqref{eq:model-distances} gives
\begin{equation}\label{eq:density-transport-exact}
 d_W(H,J)=\int_0^1|F_J(x)-F_H(x)|\,dx=Bw^2=\varepsilon.
\end{equation}
In particular, for every sample size $n$, the two environments
\begin{equation}\label{eq:density-common-history-worlds}
 (\mu,\nu_{1:n})=(H,H,\ldots,H)\quad\text{under }K,
 \qquad
 (\mu,\nu_{1:n})=(J,H,\ldots,H)\quad\text{under }W
\end{equation}
are admissible in both $F$ and $A$. Their historical joint distribution is the same product law $H^{\otimes n}$, and $H$ is independent of $n$. No IID restriction is imposed on the \emph{policy}: the IID array here is simply one environment in the supremum over all independent arrays.

To identify the Kolmogorov-safe anchor price, compute the entire historical revenue function:
\begin{equation}\label{eq:density-center-revenue}
 r_H(p)=
 \begin{cases}
 p(1-Bp),&0\le p\le d,\\
 cp,&d\le p\le a,\\
 Bp(1-p),&a\le p\le1.
 \end{cases}
\end{equation}
The first interval contains $b=1/(2B)$ and has maximum $r_H(b)=1/(4B)$. The middle expression is strictly increasing, and the final expression is strictly decreasing because $a>1/2$. Since
\begin{equation}\label{eq:density-revenue-near-tie}
 r_H(a)=ac=Ba(1-a)=\frac1{4B}+m=r_H(b)+m,
\end{equation}
$a$ is the \emph{unique} global revenue maximizer of $H$, but its advantage over the distant price $b$ is only $m=o(\sqrt\varepsilon)$. Moreover, for
\begin{equation}\label{eq:density-kappa}
 \kappa=\min\{c,B(2a-1)\}>0,
\end{equation}
we have the all-price inequalities
\begin{equation}\label{eq:density-anchor-forcing}
 r_H(a)-r_H(p)\ge
 \begin{cases}
 m,&0\le p\le d,\\
 \kappa|p-a|,&d<p\le1.
 \end{cases}
\end{equation}
For $d<p\le a$ the regret is exactly $c(a-p)$; for $p\ge a$ it equals $B(p-a)(p+a-1)\ge B(2a-1)(p-a)$. The constants $c$ and $\kappa$ converge to strictly positive limits for fixed $B\ge2$.

Let $\pi=(\pi_n)$ be \emph{any} admissible randomized policy sequence and let $Q_n$ be its unconditional output-price law when the historical observations are drawn independently from $H$. This integrates both the history and all additional policy randomization; it is the same $Q_n$ in the two worlds of \eqref{eq:density-common-history-worlds}. Put
\[
 h_n=\int\bigl(r_H(a)-r_H(p)\bigr)\,Q_n(dp).
\]
The $K$-anchor is admissible, so $0\le h_n\le R_{K,n}^E(\pi;\mathcal D_B,\varepsilon)$. Applying \eqref{eq:density-anchor-forcing} to the \emph{whole} law $Q_n$ gives
\begin{equation}\label{eq:density-price-control}
 Q_n([0,d])\le\frac{h_n}{m},\qquad
 \int_{(d,1]}|p-a|\,Q_n(dp)\le\frac{h_n}{\kappa}.
\end{equation}
There is no restriction to deterministic prices, symmetric policies, or policies using only the empirical distribution.

Define the revenue perturbation $D(p)=r_H(p)-r_J(p)$. Since both laws are atomless, \eqref{eq:density-cdf-triangle} implies
\begin{equation}\label{eq:density-perturbation}
 D(p)=p\,[F_J(p)-F_H(p)],\qquad
 D(a)=aBw,\qquad 0\le D(p)\le Bw.
\end{equation}
On each differentiable piece, $|D'(p)|\le Bw+B$, because the CDF difference is at most $Bw$ and has slope of absolute value at most $B$; continuity across the boundaries extends this to the global Lipschitz bound
\begin{equation}\label{eq:density-D-Lip}
 |D(p)-D(q)|\le(B+Bw)|p-q|\quad(p,q\in[0,1]).
\end{equation}
On $p\le d$ retain the nonnegative bound $D(p)\ge0$, and on $p>d$ use \eqref{eq:density-D-Lip} centered at $a$. Equation \eqref{eq:density-price-control} yields
\begin{equation}\label{eq:density-transfer}
 \begin{split}
 \int D(p)\,Q_n(dp)
 &\ge aBw\bigl(1-Q_n([0,d])\bigr)
       -(B+Bw)\int_{(d,1]}|p-a|\,Q_n(dp)\\
 &\ge aBw-\left(\frac{aBw}{m}+\frac{B+Bw}{\kappa}\right)h_n.
 \end{split}
\end{equation}
Because the transported block lies above $b$, the sales probability at $b$ is unchanged: $r_J(b)=r_H(b)=1/(4B)$. The $W$-future's optimal revenue is at least this value. Consequently, against the second world,
\begin{equation}\label{eq:density-w-finite-lower}
 \begin{split}
 \int\bigl(r_J^\star-r_J(p)\bigr)\,Q_n(dp)
 &\ge r_J(b)-\int r_J(p)\,Q_n(dp)\\
 &=r_H(b)-\int r_H(p)\,Q_n(dp)+\int D(p)\,Q_n(dp)\\
 &\ge -m+aBw-\left(\frac{aBw}{m}+\frac{B+Bw}{\kappa}\right)h_n.
 \end{split}
\end{equation}
Both worlds belong to the respective finite-$n$ environmental suprema, hence, writing $M_\varepsilon=aBw/m+(B+Bw)/\kappa$,
\begin{equation}\label{eq:density-all-policy-finite}
 R_{W,n}^E(\pi;\mathcal D_B,\varepsilon)
 +M_\varepsilon R_{K,n}^E(\pi;\mathcal D_B,\varepsilon)
 \ge aBw-m\qquad(n\ge1).
\end{equation}
This inequality is already uniform in $n$ and every admissible policy; in particular it does not rely on a statistical-to-population reduction. Since $M_\varepsilon\ge0$, the definition of the two sequence risks and the eventual upper bound implicit in the $K$-risk limsup imply
\begin{equation}\label{eq:density-all-policy-asymptotic}
 R_W^E(\pi;\mathcal D_B,\varepsilon)
 \ge aBw-m-M_\varepsilon R_K^E(\pi;\mathcal D_B,\varepsilon).
\end{equation}
Thus, if $R_K^E(\pi;\mathcal D_B,\varepsilon)\le L\varepsilon$,
\begin{equation}\label{eq:density-constrained-finite-radius}
 R_W^E(\pi;\mathcal D_B,\varepsilon)
 \ge aBw\left(1-\frac{L\varepsilon}{m}\right)
      -m-\frac{(B+Bw)L\varepsilon}{\kappa}.
\end{equation}
The right side depends only on $B,L,\varepsilon$, not on $\pi$, so the same inequality holds after infimizing over all $K$-feasible policy sequences. The feasible set is nonempty for $L\ge2$ by the Kolmogorov-safe empirical-revenue policy established in Section~\ref{sec:model}.

Finally, $aBw=a\sqrt{B\varepsilon}$, $m/\sqrt\varepsilon=\varepsilon^{1/4}\to0$, $L\varepsilon/m=L\varepsilon^{1/4}\to0$, and $\kappa$ is bounded away from zero. Dividing \eqref{eq:density-constrained-finite-radius} by $\sqrt\varepsilon$, then letting $\varepsilon\downarrow0$ with $B,L$ fixed, yields \eqref{eq:density-constrained-lower} because $a\to P_B$.
\end{proof}

\subsection{Optimal-price ceiling and clipped Wasserstein smoothing}\label{app:density-oracle}

\begin{lemma}[Optimal-price ceiling]\label{lem:density-price-ceiling}
For every $B\ge1$ and every $\mu\in\mathcal D_B$, every optimal posted price lies in $[0,P_B]$. The upper endpoint $P_B$ cannot be uniformly reduced over $\mathcal D_B$ when $B>1$.
\end{lemma}
\begin{proof}
For $p_0=1/(2B)$, the density bound gives $F_\mu(p_0)\le Bp_0=1/2$, so, using atomlessness,
\begin{equation}\label{eq:density-price-ceiling-lower}
 r_\mu^\star\ge r_\mu(p_0)
 =p_0[1-F_\mu(p_0)]\ge\frac1{4B}.
\end{equation}
Conversely, $\mu([p,1])\le B(1-p)$ and therefore
\begin{equation}\label{eq:density-price-ceiling-upper}
 r_\mu(p)\le Bp(1-p).
\end{equation}
Since $P_B$ is the larger root of $p(1-p)=1/(4B^2)$, the last expression is strictly smaller than $1/(4B)$ for $p>P_B$. No such price is optimal.

To see sharpness of this \emph{price ceiling} (not of any minimax coefficient), let $d_0=P_B+1/B-1$ and consider
\[
 f_0(x)=B\mathbf1_{[0,d_0]}(x)+B\mathbf1_{[P_B,1]}(x).
\]
For $B>1$, $0<1/(2B)\le d_0<P_B$ because $d_0-1/(2B)=[\sqrt{B^2-1}-(B-1)]/(2B)\ge0$; its total mass is $B(d_0+1-P_B)=1$. Its revenue is $p(1-Bp)$ on $[0,d_0]$, $Bp(1-P_B)$ on $[d_0,P_B]$, and $Bp(1-p)$ on $[P_B,1]$. The first expression reaches $1/(4B)$ at $1/(2B)$, the second increases to $BP_B(1-P_B)=1/(4B)$, and the third decreases because $P_B>1/2$. Thus $P_B$ itself is an optimal price for this admissible law.
\end{proof}

\begin{lemma}[Density-specific clipped smoothing oracle]\label{lem:density-clipped-oracle}
For fixed $B\ge1$, $E\in\{F,A\}$, and $0<\varepsilon\le P_B/2$, there is a measurable sample-dependent randomized pricing policy $\pi^{\mathrm{clip}}$ such that, at every $n\ge1$,
\begin{equation}\label{eq:density-clipped-finite}
 R_{W,n}^E(\pi^{\mathrm{clip}};\mathcal D_B,\varepsilon)
 \le\sqrt{2P_B\varepsilon}
 +\frac{P_B}{2\sqrt{2\varepsilon/P_B}\,\sqrt n}.
\end{equation}
In particular,
\begin{equation}\label{eq:density-oracle-upper}
 \limsup_{\varepsilon\downarrow0}
 \frac{W_E^\star(\mathcal D_B,\varepsilon)}{\sqrt\varepsilon}
 \le U_B=\sqrt{2P_B}.
\end{equation}
\end{lemma}
\begin{proof}
Abbreviate $P=P_B$ and define the nondecreasing, $1/P$-Lipschitz clipping/rescaling map
\begin{equation}\label{eq:density-clip-map}
 T:[0,1]\to[0,1],\qquad T(v)=\frac{\min\{v,P\}}P.
\end{equation}
For a law $\eta$, let $\widetilde\eta=T_\#\eta$ be its pushforward. For every $q\in[0,1]$, including $q=0$ and $q=1$,
\begin{equation}\label{eq:density-revenue-scaling}
 r_{\widetilde\mu}(q)
 =q\Pr_\mu(T(V)\ge q)
 =q\Pr_\mu(V\ge Pq)
 =P^{-1}r_\mu(Pq).
\end{equation}
Lemma~\ref{lem:density-price-ceiling} ensures that $\mu$ has a revenue optimizer in $[0,P]$, whence
\begin{equation}\label{eq:density-optimal-regret-scaling}
 r_{\widetilde\mu}^\star=P^{-1}r_\mu^\star,
 \qquad
 \ell_\mu(\mathcal L(PQ))=P\,\ell_{\widetilde\mu}(\mathcal L(Q))
\end{equation}
for any randomized transformed price $Q\in[0,1]$; $\mathcal L$ denotes a probability law. There is no additive pricing intercept in the transformation, so this is an \emph{exact} revenue identity, not merely a comparison of nearby price intervals.

For any pair of valuation laws $\mu,\eta$, the coupling characterization of $W_1$ and the Lipschitz property of $T$ imply
\begin{equation}\label{eq:density-clipped-transport}
 d_W(\widetilde\mu,\widetilde\eta)
 \le P^{-1}d_W(\mu,\eta).
\end{equation}
Apply $T$ separately to every independent historical observation $X_i\sim\nu_i$. The transformed observations remain independent, with possibly different marginals $\widetilde\nu_i$, and each satisfies $d_W(\widetilde\nu_i,\widetilde\mu)\le\varepsilon/P$. Under experiment $A$, both transformed future and historical laws belong to the image class $T_\#\mathcal D_B$; under $F$ only the transformed future is restricted to this class. Both transformed environments are therefore covered by Lemma~\ref{lem:oracle-universal} with class $T_\#\mathcal D_B$ and radius $\delta=\varepsilon/P$. The image laws may have atoms at $1$; this is why it is essential that the approved universal smoothing lemma does not require bounded densities.

For completeness, the resulting \emph{density-specific} policy can be written explicitly. Given $X_{1:n}$, set $Y_i=T(X_i)$ and $\widehat\rho_n=n^{-1}\sum_i\delta_{Y_i}$; take $t=\sqrt{2\delta}=\sqrt{2\varepsilon/P}\le1$. Choose the smallest maximizer
\begin{equation}\label{eq:density-clipped-policy}
 \widehat z_n\in\arg\max_{z\in[t,1]}
 \frac1t\int_{z-t}^z r_{\widehat\rho_n}(q)\,dq,
 \qquad Q\sim\operatorname{Unif}[\widehat z_n-t,\widehat z_n],
 \qquad p=PQ.
\end{equation}
The continuous moving-integral objective admits a measurable smallest maximizer as proved in Lemma~\ref{lem:oracle-universal}; the final uniform draw makes the policy a valid stochastic kernel. This policy is evaluated against all Wasserstein-admissible independent historical arrays, not a single population center. In the transformed experiment, the cited lemma proves uniformly that
\[
 R_{W,n}^{\widetilde E}\le\sqrt{2\delta}
       +\frac1{2\sqrt{2\delta}\,\sqrt n};
\]
its proof uses the $1/t$ \emph{pairwise} smoothed-revenue Lipschitz inequality and $\mathbb E d_W(\widehat\rho_n,n^{-1}\sum_i\widetilde\nu_i)\le1/(2\sqrt n)$. Multiplying this finite-$n$ risk by $P$ through \eqref{eq:density-optimal-regret-scaling} gives \eqref{eq:density-clipped-finite}. Holding $\varepsilon>0$ fixed, take $n\to\infty$ first to obtain
\[
 W_E^\star(\mathcal D_B,\varepsilon)\le\sqrt{2P\varepsilon}.
\]
The normalized bound \eqref{eq:density-oracle-upper} follows immediately. Note that this is the sharper $\sqrt{2P_B}$ coefficient; applying the unscaled universal bound directly would only give $\sqrt2$.
\end{proof}

\subsection{Strict difference, ratio, and the certificate boundary}\label{app:density-comparison}

\begin{proof}[Proof of Theorem~\ref{thm:density-gap}]
For fixed $B>17/8>2$, Lemma~\ref{lem:density-two-peak} and Lemma~\ref{lem:density-clipped-oracle} establish \eqref{eq:density-global-certificates} in both experiments, with the prescribed statistical and small-radius order of limits. Since $\Delta_L^E=A_L^E-W_E^\star$ at each radius, the elementary liminf--limsup inequality gives
\begin{equation}\label{eq:density-difference-derivation}
 \liminf_{\varepsilon\downarrow0}
 \frac{\Delta_L^E}{\sqrt\varepsilon}
 \ge\liminf_{\varepsilon\downarrow0}\frac{A_L^E}{\sqrt\varepsilon}
     -\limsup_{\varepsilon\downarrow0}\frac{W_E^\star}{\sqrt\varepsilon}
 \ge C_B-U_B,
\end{equation}
where the common arguments $(\mathcal D_B,\varepsilon)$ are suppressed only in this display.

We verify positivity of the denominator in the ratio, without assuming a sharp Wasserstein-only lower coefficient. Reuse $H,J$ from \eqref{eq:density-two-peak-laws}. By \eqref{eq:density-revenue-near-tie} and \eqref{eq:density-perturbation},
\[
 r_J(b)=\frac1{4B},\qquad
 r_J(a)=\frac1{4B}+m-aBw.
\]
For all sufficiently small $\varepsilon$, $aBw>m$ because $aBw/m=a\sqrt B\,\varepsilon^{-1/4}\to\infty$. Thus $a$, the unique revenue maximizer of $H$, is \emph{not} a revenue maximizer of $J$. Continuity of the two revenue functions on the compact price interval gives a strictly positive constant (for each fixed $\varepsilon$)
\begin{equation}\label{eq:density-positive-denominator}
 \gamma_\varepsilon=\min_{p\in[0,1]}
 \bigl\{r_H^\star-r_H(p)+r_J^\star-r_J(p)\bigr\}>0.
\end{equation}
Both $(H,H,\ldots,H)$ and $(J,H,\ldots,H)$ are Wasserstein-admissible in $F$ and $A$ by \eqref{eq:density-transport-exact}. Their common historical joint law makes the same (possibly randomized) price distribution apply in both worlds. For every $n$ and every policy, their two expected regrets sum to at least $\gamma_\varepsilon$, so the larger is at least $\gamma_\varepsilon/2$. Consequently,
\begin{equation}\label{eq:density-denominator-lower}
 W_E^\star(\mathcal D_B,\varepsilon)\ge\gamma_\varepsilon/2>0
\end{equation}
for all sufficiently small positive radii. This ensures the ratio in the theorem is defined; it does \emph{not} identify the Wasserstein-only minimax coefficient.

Now for any $\eta>0$, sufficiently small $\varepsilon$ satisfy
\[
 A_L^E(\mathcal D_B,\varepsilon)\ge(C_B-\eta)\sqrt\varepsilon,
 \quad
 0<W_E^\star(\mathcal D_B,\varepsilon)\le(U_B+\eta)\sqrt\varepsilon.
\]
Choosing $\eta<C_B$ and dividing, then letting $\eta\downarrow0$, yields
\begin{equation}\label{eq:density-ratio-derivation}
 \liminf_{\varepsilon\downarrow0}
 \frac{A_L^E(\mathcal D_B,\varepsilon)}{W_E^\star(\mathcal D_B,\varepsilon)}
 \ge\frac{C_B}{U_B}.
\end{equation}

Finally, since $P_B>0$, direct algebra gives
\begin{equation}\label{eq:density-boundary-algebra}
 C_B>U_B
 \iff BP_B>2
 \iff B+\sqrt{B^2-1}>4.
\end{equation}
The function $B\mapsto B+\sqrt{B^2-1}$ is strictly increasing for $B>1$, and at $B=17/8$ it equals $17/8+15/8=4$; equivalently $P_{17/8}=16/17$. Thus
\[
 C_B-U_B>0\quad\text{and}\quad C_B/U_B>1
 \quad\Longleftrightarrow\quad B>17/8
\]
within the proven coefficient comparison. This is a sufficient range for a strict global minimax gap, \emph{not} a sharp phase-transition assertion about the underlying problem. In particular the comparison does not resolve $\mathcal D_2$. Equations \eqref{eq:density-difference-derivation} and \eqref{eq:density-ratio-derivation} complete the proof.
\end{proof}

\section{Proofs for Constructive Finite-Sample Pricing}\label{app:finite-sample}

This appendix proves Proposition~\ref{prop:fs-simultaneous}, including its implementation and budget certificate. Fix $n\ge1$, a nonempty class $\mathcal C\subseteq\mathcal P$, and $\varepsilon\ge0$. All results apply separately to experiments $F$ and $A$ of \eqref{eq:model-environments}; no convexity or regularity of $\mathcal C$ is assumed. Each historical array is independent but may be nonidentically distributed, and full valuations are observed.

\subsection{Measurable empirical optimization and non-IID concentration}\label{app:fs-empirical}

For arbitrary historical marginals $\nu_1,\ldots,\nu_n\in\mathcal P$, write
\begin{equation}\label{eq:fs-historical-laws}
 \bar\nu_n=\frac1n\sum_{i=1}^n\nu_i,\qquad
 \widehat\nu_n=\frac1n\sum_{i=1}^n\delta_{X_i},\qquad
 \widehat r_n(p)=r_{\widehat\nu_n}(p),\qquad
 \eta_n=\sup_{p\in[0,1]}\bigl|\widehat r_n(p)-r_{\bar\nu_n}(p)\bigr|.
\end{equation}
For any probability law $\rho$ on $[0,1]$, $r_\rho(p)=p\rho([p,1])$ is left-continuous for $p\in(0,1]$, and $r_\rho(0)=0$. Hence $\eta_n$ is the supremum of the same absolute difference on the countable set $(\mathbb Q\cap(0,1))\cup\{1\}$ and is measurable. For a realized sample, $\widehat r_n$ increases linearly between distinct positive sample valuations and can decrease only immediately above a valuation. Its maximum is therefore attained in $\{0,X_1,\ldots,X_n\}$. Choosing the smallest maximizer as in \eqref{eq:fs-empirical-rule} is Borel measurable: it is a finite comparison of measurable functions with a fixed tie-break. The map $x\mapsto\mathcal L((\widehat p_n(x)-U)_+)$ defines a Borel stochastic kernel, since for every Borel $D\subseteq[0,1]$,
\[
 \pi_{n,t}(D\mid x)=\frac1t\int_0^t\mathbf 1\{(\widehat p_n(x)-u)_+\in D\}\,du
\]
is measurable by joint measurability and Tonelli's theorem. Sorting the sample and computing empirical tail counts requires $O(n\log n)$ operations.

\begin{lemma}[Nonidentical-sample empirical-process control]\label{lem:fs-empirical}
For every independent array $X_i\sim\nu_i$ on $[0,1]$,
\begin{equation}\label{eq:fs-empirical-expected}
 \mathbb E_{\nu_{1:n}}\eta_n\le\frac4{\sqrt n}.
\end{equation}
If $\xi_n=d_K(\widehat\nu_n,\bar\nu_n)$, then also $\mathbb E\xi_n\le4/\sqrt n$. For every $0<\delta<1$, with
\begin{equation}\label{eq:fs-deviation-radius}
 z_n(\delta)=\frac4{\sqrt n}+\sqrt{\frac{\log(1/\delta)}{2n}},
\end{equation}
we have separately
\begin{equation}\label{eq:fs-concentration}
 \Pr\{\eta_n>z_n(\delta)\}\le\delta,
 \qquad \Pr\{\xi_n>z_n(\delta)\}\le\delta.
\end{equation}
Both inequalities hold simultaneously with probability at least $1-\delta$ if $z_n(\delta)$ is replaced in both events by $z_n(\delta/2)$.
\end{lemma}
\begin{proof}
For $p\in[0,1]$ define $f_p(v)=p\mathbf 1\{v\ge p\}$, which takes values in $[0,1]$. Introduce independent ghost observations $X'_i\sim\nu_i$, independent across $i$ and of $X_{1:n}$, and independent Rademacher signs $\sigma_1,\ldots,\sigma_n$. Symmetrization with a separate ghost law at each coordinate gives
\begin{equation}\label{eq:fs-symmetrization}
 \begin{split}
 \mathbb E\eta_n
 &\le\mathbb E\sup_p\left|\frac1n\sum_{i=1}^n\bigl(f_p(X_i)-f_p(X'_i)\bigr)\right|\\
 &\le\frac2n\mathbb E_{X,\sigma}\sup_p\left|\sum_{i=1}^n\sigma_i f_p(X_i)\right|.
 \end{split}
\end{equation}
The first step uses Jensen's inequality conditional on $X_{1:n}$, since $\mathbb E f_p(X'_i)=\int f_p\,d\nu_i$. The second uses invariance of each independent pair $(X_i,X'_i)$ under exchange, followed by the triangle inequality; identical distribution \emph{between coordinates} is not needed.

Condition on $X_{1:n}$ and order the observations by valuation, breaking valuation ties by their original indices. Every set $\{i:X_i\ge p\}$ is a suffix of this order (some suffix lengths may be unavailable because of ties). Since $0\le p\le1$,
\begin{equation}\label{eq:fs-suffix-bound}
 \sup_p\left|\sum_{i=1}^n\sigma_i f_p(X_i)\right|
 \le\max_{0\le k\le n}\left|\sum_{j=1}^k\sigma_{(j)}\right|.
\end{equation}
Conditional on the sample, the reordered signs remain independent Rademacher variables. Their partial sums form an $L^2$ martingale, so Doob's maximal inequality and Cauchy--Schwarz give
\[
 \mathbb E_\sigma\max_{0\le k\le n}\left|\sum_{j=1}^k\sigma_{(j)}\right|
 \le\left(4\mathbb E_\sigma\left|\sum_{j=1}^n\sigma_{(j)}\right|^2\right)^{1/2}
 =2\sqrt n.
\]
Substitution into \eqref{eq:fs-symmetrization} proves \eqref{eq:fs-empirical-expected}. Applying the same argument to the threshold functions $v\mapsto\mathbf 1\{v\le x\}$, whose selected sets are prefixes, proves the estimate for $\xi_n$ (the supremum in $d_K$ is measurable because CDFs are right-continuous).

Replacing a single $X_i$ changes $\widehat r_n(p)$ at any price by at most $1/n$, uniformly in $p$, and therefore changes $\eta_n$ by at most $1/n$. The same bounded-differences property holds for $\xi_n$. McDiarmid's inequality, valid for independent nonidentical variables, gives for either statistic $Z\in\{\eta_n,\xi_n\}$ and $a>0$,
\[
 \Pr\{Z-\mathbb EZ>a\}\le\exp(-2na^2).
\]
Use the expectation bounds and $a=\sqrt{\log(1/\delta)/(2n)}$ to obtain \eqref{eq:fs-concentration}; the last claim follows by a union bound.
\end{proof}

For every realized sample the empirical maximizer satisfies
\begin{equation}\label{eq:fs-erm-comparison}
 r_{\bar\nu_n}(\widehat p_n)\ge r_{\bar\nu_n}^\star-2\eta_n.
\end{equation}
Indeed, if $p^\star\in\arg\max r_{\bar\nu_n}$, empirical optimality at $\widehat p_n$ implies
\[
 r_{\bar\nu_n}(p^\star)-r_{\bar\nu_n}(\widehat p_n)
 \le\bigl(r_{\bar\nu_n}(p^\star)-\widehat r_n(p^\star)\bigr)
 +\bigl(\widehat r_n(\widehat p_n)-r_{\bar\nu_n}(\widehat p_n)\bigr)
 \le2\eta_n.
\]
An optimum $p^\star$ exists because $p\mapsto r_\rho(p)$ is upper semicontinuous on the compact price interval: it is left-continuous and its right limit is no larger than its value (this includes the purchase-at-equality convention).

\subsection{One-sided price comparisons}\label{app:fs-one-sided}

For either metric $d\in\{d_K,d_W\}$, its one-dimensional CDF representation and the triangle inequality imply
\begin{equation}\label{eq:fs-barycenter-radius}
 d(\mu,\bar\nu_n)\le\frac1n\sum_{i=1}^n d(\mu,\nu_i)\le\varepsilon
\end{equation}
whenever the array is $d$-admissible. For $d_K$, use the supremum of the average CDF differences; for $d_W$, integrate their absolute average over $[0,1]$. This implication does not require $\bar\nu_n\in\mathcal C$, and hence applies under either $F$ or $A$.

\begin{lemma}[Downward undercutting comparisons]\label{lem:fs-one-sided}
Let $\mu,\nu\in\mathcal P$, $q\in[0,1]$, $t>0$, and $U\sim\operatorname{Unif}[0,t]$ independent of valuations. Then
\begin{align}
 r_\nu((q-u)_+)&\ge r_\nu(q)-u &&(u\ge0),\label{eq:fs-price-monotonicity}\\
 \left|r_\mu(q)-r_\nu(q)\right|&\le d_K(\mu,\nu),\label{eq:fs-k-price}\\
 \mathbb E_U r_\mu((q-U)_+)&\ge r_\nu(q)-t-\frac{d_W(\mu,\nu)}t.\label{eq:fs-w-one-sided}
\end{align}
Consequently,
\begin{equation}\label{eq:fs-optima-comparison}
 r_\mu^\star\le r_\nu^\star+d_K(\mu,\nu),\qquad
 r_\mu^\star\le r_\nu^\star+2\sqrt{d_W(\mu,\nu)}.
\end{equation}
The Wasserstein comparison remains valid when $d_W(\mu,\nu)=0$, by the zero-distance identity.
\end{lemma}
\begin{proof}
Writing $q_u=(q-u)_+$, the purchase set at $q_u$ contains the purchase set at $q$, and $q_u\ge q-u$. Thus
\[
 r_\nu(q_u)\ge q_u\nu([q,1])\ge r_\nu(q)-u,
\]
which proves \eqref{eq:fs-price-monotonicity}, including $q<u$ and atoms at the price. For $q>0$, the survival probability is $\rho([q,1])=1-F_\rho(q-)$; left limits of CDFs inherit their uniform Kolmogorov bound. Multiplication by $q\le1$ gives \eqref{eq:fs-k-price}; the case $q=0$ is immediate.

For \eqref{eq:fs-w-one-sided}, couple $X\sim\nu$ and $Y\sim\mu$ with any coupling $\gamma$, independently of $U$. If $X\ge q$ and $Y<q_U$, then $q_U>0$ and
\[
 U<q-Y\le X-Y\le|X-Y|.
\]
For each coupled pair, the chance of this event over $U$ is therefore at most $\min\{1,|X-Y|/t\}\le|X-Y|/t$. Pointwise, since $0\le q-q_U\le t$ and $q_U\le1$,
\[
 q\mathbf 1\{X\ge q\}-q_U\mathbf 1\{Y\ge q_U\}
 \le (q-q_U)\mathbf 1\{X\ge q\}
      +\mathbf 1\{X\ge q,\,Y<q_U\}.
\]
Take expectation over $U$ and $\gamma$, then infimize over couplings (or use an optimal coupling, which exists on $[0,1]$) to obtain \eqref{eq:fs-w-one-sided}.

The first optimal-value comparison follows by maximizing \eqref{eq:fs-k-price}. For the second, interchange $\mu$ and $\nu$ in \eqref{eq:fs-w-one-sided}: for every $s>0$ and every $q$,
\[
 r_\mu(q)\le\mathbb E_{U_s}r_\nu((q-U_s)_+)+s+\frac{d_W(\mu,\nu)}s
 \le r_\nu^\star+s+\frac{d_W(\mu,\nu)}s.
\]
Maximize over $q$ and choose $s=\sqrt{d_W(\mu,\nu)}$ when the distance is positive. If it is zero, the laws coincide, and the optima are equal.
\end{proof}

\subsection{Expected regret, conditional risk, and confidence events}\label{app:fs-regret}

\begin{proof}[Proof of Proposition~\ref{prop:fs-simultaneous}]
Fix a metric-$d$-admissible environment $(\mu,\nu_{1:n})$. The bound \eqref{eq:fs-barycenter-radius} gives $d(\mu,\bar\nu_n)\le\varepsilon$. Conditional on a realized sample $x_{1:n}$, average only over the independent policy draw $U$. In the Kolmogorov world, \eqref{eq:fs-optima-comparison}, \eqref{eq:fs-k-price}, \eqref{eq:fs-price-monotonicity}, and \eqref{eq:fs-erm-comparison} yield
\begin{equation}\label{eq:fs-conditional-k}
 \begin{split}
 r_\mu^\star-\mathbb E_Ur_\mu((\widehat p_n-U)_+)
 &\le r_{\bar\nu_n}^\star+\varepsilon
       -\mathbb E_Ur_{\bar\nu_n}((\widehat p_n-U)_+)+\varepsilon\\
 &\le r_{\bar\nu_n}^\star-r_{\bar\nu_n}(\widehat p_n)+2\varepsilon+t
 \le 2\varepsilon+t+2\eta_n.
 \end{split}
\end{equation}
In the Wasserstein world, \eqref{eq:fs-optima-comparison}, \eqref{eq:fs-w-one-sided}, and \eqref{eq:fs-erm-comparison} give
\begin{equation}\label{eq:fs-conditional-w}
 \begin{split}
 r_\mu^\star-\mathbb E_Ur_\mu((\widehat p_n-U)_+)
 &\le r_{\bar\nu_n}^\star+2\sqrt\varepsilon
       -r_{\bar\nu_n}(\widehat p_n)+t+\frac\varepsilon t\\
 &\le 2\sqrt\varepsilon+t+\frac\varepsilon t+2\eta_n.
 \end{split}
\end{equation}
These are \emph{samplewise inequalities for conditional expected regret}, not assertions for every realized policy coin $U$. Now take expectation with respect to the independent, potentially nonidentical sample, apply \eqref{eq:fs-empirical-expected}, and only then take the supremum over the corresponding environment set in \eqref{eq:model-finite-risk}. The resulting coefficients $2\mathbb E\eta_n\le8/\sqrt n$ are uniform across arrays, and prove both inequalities in \eqref{eq:fs-main-bounds}. The policy is measurable by the argument preceding Lemma~\ref{lem:fs-empirical}; the construction is identical for $F$ and $A$ and for the two metrics.
\end{proof}

For completeness, the distinction between the three risk statements is explicit. For each \emph{fixed} admissible environment, Lemma~\ref{lem:fs-empirical} and \eqref{eq:fs-conditional-k}--\eqref{eq:fs-conditional-w} imply, with probability at least $1-\delta$ over $X_{1:n}$,
\begin{equation}\label{eq:fs-high-prob-conditional}
 \begin{aligned}
 r_\mu^\star-\mathbb E_U[r_\mu(P_{n,t})\mid X_{1:n}]
 &\le2\varepsilon+t+2z_n(\delta) &&\text{under }K,\\
 r_\mu^\star-\mathbb E_U[r_\mu(P_{n,t})\mid X_{1:n}]
 &\le2\sqrt\varepsilon+t+\varepsilon/t+2z_n(\delta) &&\text{under }W.
 \end{aligned}
\end{equation}
The $K$ and $W$ displays refer to their \emph{respective} admissible environments; they are not a single joint probability statement for two different sample distributions. For a fixed historical array, each event $\{\eta_n\le z_n(\delta)\}$ supports the corresponding bound uniformly over all future $\mu$ satisfying the required distance to $\bar\nu_n$. By contrast, $R_{d,n}^E$ is a deterministic supremum of \emph{unconditional expectations}, and the high-probability event alone neither establishes a single-realization regret bound nor changes the order of the supremum and expectation. As all regrets lie in $[0,1]$, a conditional event of probability $1-\delta$ with threshold $a$ would by itself yield only $\mathbb E\ell\le a+\delta$; the sharper expected certificates above follow directly from the expectation bound on $\eta_n$.

\subsection{Certified budget widths and a qualified statistical floor}\label{app:fs-budget}

\begin{lemma}[Budget feasibility of the reported certificate]\label{lem:fs-budget}
Fix $\varepsilon>0$, $n\ge1$, and a prescribed finite-sample Kolmogorov expected-regret budget $b$. Within the conservative certificates \eqref{eq:fs-main-bounds}, a strictly positive undercutting width is certified if and only if
\begin{equation}\label{eq:fs-positive-slack}
 s_n:=b-2\varepsilon-\frac8{\sqrt n}>0.
\end{equation}
When this holds, all $0<t\le s_n$ are certified and the width \eqref{eq:fs-width-choice} minimizes the reported Wasserstein upper bound among them. If $b>2\varepsilon$, the sufficient sample condition is exactly $n>64/(b-2\varepsilon)^2$. These are statements about the \emph{given upper certificates}, not necessary conditions for existence of a safe policy. At $\varepsilon=0$ and $s_n>0$, the same certificates allow every $0<t\le s_n$, and the infimum of $t+\varepsilon/t=t$ is approached as $t\downarrow0$, not attained at a positive width.
\end{lemma}
\begin{proof}
The first line of \eqref{eq:fs-main-bounds} certifies safety precisely when $2\varepsilon+t+8/\sqrt n\le b$, or $0<t\le s_n$. Its feasible interval is nonempty exactly when $s_n>0$. For $\varepsilon>0$, the function $h(t)=t+\varepsilon/t$ has derivative $1-\varepsilon/t^2$, and is strictly decreasing on $(0,\sqrt\varepsilon]$ and increasing on $[\sqrt\varepsilon,\infty)$. Hence its constrained minimizer is $\min\{\sqrt\varepsilon,s_n\}$. Rearranging $8/\sqrt n<b-2\varepsilon$ with $b>2\varepsilon$ yields the sample-size condition. The zero-radius claim is immediate. When $s_n\le0$, the certificate says nothing about alternative policy families or tighter statistical estimates.
\end{proof}

The original finite-sample source also gives a lower floor, but only for a specified continuous bounded-density class. The following verification preserves that restriction; it is \emph{not} a lower bound for arbitrary $\mathcal C$ or a sharp finite-sample frontier.

\begin{lemma}[Two-point statistical lower floor on $\mathcal D_B$]\label{lem:fs-finite-floor}
For every $B\ge12$, $n\ge1$, $\varepsilon\ge0$, $E\in\{F,A\}$, $d\in\{K,W\}$, and every measurable randomized, sample-dependent pricing kernel $\pi_n$,
\begin{equation}\label{eq:fs-statistical-floor}
 R_{d,n}^E(\pi;\mathcal D_B,\varepsilon)\ge\frac1{50\sqrt n}.
\end{equation}
In particular, a prescribed $K$ budget $b<1/(50\sqrt n)$ is infeasible on this class, despite allowing zero heterogeneity. This is independent of, and not a converse to, Lemma~\ref{lem:fs-budget}.
\end{lemma}
\begin{proof}
Set $\delta=1/(16\sqrt n)$, $w_\pm=1/2\pm\delta$, and define
\begin{equation}\label{eq:fs-two-point-laws}
 \mu_\pm=(1-w_\pm)\operatorname{Unif}[9/20,11/20]
                  +w_\pm\operatorname{Unif}[9/10,1].
\end{equation}
Since $0<\delta\le1/16$, each component weight lies in $[7/16,9/16]$. Both intervals have length $1/10$, so each law has density at most $10(9/16)=45/8<12\le B$ and belongs to $\mathcal D_B$. At price $p_L=9/20$, each law earns $9/20$; at price $p_H=9/10$, revenue is $(9/10)w_\pm=9/20\pm(9/10)\delta$. For $p\le p_L$, revenue is at most the revenue at $p_L$. Within $[p_L,11/20]$,
\[
 r_{\mu_\pm}(p)=p\left[w_\pm+(1-w_\pm)\frac{11/20-p}{1/10}\right],
 \quad r'_{\mu_\pm}(p)=w_\pm+(1-w_\pm)(11/2-20p)<0,
\]
because at $p=p_L$ the derivative is $w_\pm-(7/2)(1-w_\pm)<0$ for $w_\pm\le9/16$, and the derivative decreases in $p$. For $11/20<p\le p_H$, revenue is $pw_\pm$, increasing in $p$; on $[p_H,1]$, it is $10w_\pm p(1-p)$, decreasing in $p$. Thus each price $p\le11/20$ is weakly dominated by $p_L$ in \emph{both} laws, and each price $p>11/20$ by $p_H$ in both laws. The optimal price is $p_H$ for $\mu_+$ and $p_L$ for $\mu_-$, with loss $\Delta=(9/10)\delta$ for choosing the wrong one. Mapping every policy output to the dominating endpoint yields a (possibly randomized) binary testing rule without increasing revenue under either hypothesis. Therefore, writing $e_+$ and $e_-$ for its error probabilities under the respective data laws,
\begin{equation}\label{eq:fs-testing-risk}
 \max_{\pm}\mathbb E_{\mu_\pm^{\otimes n}}\ell_{\mu_\pm}(\pi_n(\cdot\mid X_{1:n}))
 \ge \Delta\max\{e_+,e_-\}
 \ge\frac\Delta2\bigl(1-\operatorname{TV}(\mu_+^{\otimes n},\mu_-^{\otimes n})\bigr).
\end{equation}
The interval containing an observation identifies a Bernoulli indicator with success probability $w_\pm$, and the conditional location distribution within each interval is the same under both laws. Hence the sample Kullback--Leibler divergence equals the Bernoulli-product divergence:
\begin{equation}\label{eq:fs-kl-calculation}
 \begin{split}
 D_{\mathrm{KL}}(\mu_+^{\otimes n}\|\mu_-^{\otimes n})
 &=2n\delta\log\frac{1+2\delta}{1-2\delta}\\
 &\le\frac{8n\delta^2}{1-4\delta^2}
 \le\frac{2}{63}<\frac1{28}.
 \end{split}
\end{equation}
The logarithm inequality follows from $\log((1+x)/(1-x))\le2x/(1-x^2)$ for $0\le x<1$; the last estimate uses $n\ge1$ and $\delta^2=1/(256n)$. Pinsker's inequality gives
$\operatorname{TV}(\mu_+^{\otimes n},\mu_-^{\otimes n})\le1/\sqrt{56}$. Consequently \eqref{eq:fs-testing-risk} is at least
\[
 \frac{9}{320\sqrt n}\left(1-\frac1{\sqrt{56}}\right)>\frac1{50\sqrt n}.
\]
Finally, in the $+$ and $-$ worlds take $\nu_i=\mu_\pm$ for every $i$: both are independent IID subexperiments of the arbitrary non-IID model, with $d(\nu_i,\mu_\pm)=0\le\varepsilon$, and are admissible under \emph{both} $F$ and $A$ for either metric. Since the maximum of these two risks is a lower bound on the full environmental supremum, \eqref{eq:fs-statistical-floor} follows for any randomized data-dependent policy.
\end{proof}

\section{Proofs for the Exact Finite-Class Frontier}\label{app:fc-proof}

This appendix proves Theorem~\ref{thm:fc-exact} for the fixed class and radius in \eqref{eq:fc-laws}. In particular, the lower certificates apply to arbitrary randomized, observation-order-dependent policies, and the matching upper construction works uniformly over all independent nonidentical historical arrays, under both experiments in \eqref{eq:model-environments}. We use $\ell_\mu(q)$ as defined in \eqref{eq:model-loss}, and temporarily suppress $\mathcal C_\mathrm{fc}$ and $\varepsilon=1/100$ in the risk notation.

\subsection{Revenue, full-price support, and metric distances}\label{app:fc-support}

Set
\[
 t_0=\frac{23}{50},\quad t_1=\frac12,\quad
 t_2=\frac{19}{30},\quad t_3=\frac{97}{150},
 \qquad 0<t_0<t_1<t_2<t_3<1.
\]
Every future law puts mass $1/4$ on its lower atom and $3/4$ on its upper atom. At the four union-support prices, direct evaluation of $r_\mu(p)=p\mu([p,1])$ gives
\begin{equation}\label{eq:fc-revenue-table-proof}
\begin{array}{c|ccc}
 p&r_H(p)&r_U(p)&r_V(p)\\ \hline
 t_0&23/50&23/50&23/50\\
 t_1&1/2&3/8&1/2\\
 t_2&19/40&19/40&19/40\\
 t_3&97/200&97/200&0
\end{array}.
\end{equation}
Purchase at equality is essential for the $t_1,t_2,t_3$ rows. Since the revenue is increasing in $p$ on each interval on which all three survival probabilities are constant, the full continuous price interval is componentwise revenue-dominated as follows: prices in $[0,t_0]$ are dominated by $t_0$; those in $(t_0,t_1]$ by $t_1$; those in $(t_1,t_2]$ by $t_2$; those in $(t_2,t_3]$ by $t_3$; and prices above $t_3$ (zero sales) by $t_2$. Moreover, $t_0$ itself is strictly dominated for all three laws by $t_2$, because $19/40>23/50$. These assertions cover the boundary prices, rather than silently passing over atoms. They also define a Borel price projection onto $\{t_1,t_2,t_3\}$ which does not increase regret for any future. The optimal revenues are
\begin{equation}\label{eq:fc-revenue-optima}
 r_H^\star=\frac12,\qquad r_U^\star=\frac{97}{200},\qquad r_V^\star=\frac12,
\end{equation}
and subtracting \eqref{eq:fc-revenue-table-proof} proves the exact matrix \eqref{eq:fc-regret-matrix}. In particular $\ell_V(p)\ge\ell_H(p)$ for every relevant price, and hence for every price by the explicit revenue formulas or the same interval calculation.

The lower atom moves from $1/2$ to $23/50$ by $1/25$ with mass $1/4$, and the upper atom moves from $97/150$ to $19/30$ by $1/75$ with mass $3/4$. The corresponding CDF differences occupy the disjoint intervals
\[
 I_1=(23/50,1/2),\quad |I_1|=1/25,
 \qquad I_2=(19/30,97/150),\quad |I_2|=1/75,
\]
with absolute heights $1/4$ and $3/4$ between $U$ and $V$. Therefore
\[
 d_W(H,U)=\frac14\frac1{25}=\frac1{100},\qquad
 d_W(H,V)=\frac34\frac1{75}=\frac1{100},\qquad
 d_W(U,V)=\frac1{50}.
\]
Taking the maximal CDF difference proves the $d_K$ row of \eqref{eq:fc-metric-table}; the overlap of the two displaced intervals does not need to be postulated (they are disjoint).

\subsection{F/A historical fibers and a valid A-adversary argument}\label{app:fc-fibers}

For any admissible deterministic array $\nu_1,\ldots,\nu_n$, let $\bar\nu_n=n^{-1}\sum_i\nu_i$. Convexity of $d_K$ and $d_W$ in their first argument gives
\begin{equation}\label{eq:fc-average-radius}
 d(\bar\nu_n,\mu)\le\frac1n\sum_{i=1}^n d(\nu_i,\mu)\le\varepsilon.
\end{equation}
In $F$, any $Q\in B_d(\mu,\varepsilon)$ is an exactly attainable average by taking every $\nu_i=Q$. Thus the $F$ average-information set is exactly $B_d(\mu,\varepsilon)$. In $A$, each $\nu_i$ lies in $\mathcal C_\mathrm{fc}\cap B_d(\mu,\varepsilon)$, so each average is a finite convex combination of that set. Conversely, any finite convex combination can be approximated by equal-weight averages with integer component counts as $n\to\infty$ (indeed in total variation, because the set is finite). This proves \eqref{eq:fc-fibers} for our finite class; no claim that a generic nonconvex class contains its barycenters is needed.

Here the admissible component sets and their barycenter sets are particularly transparent:
\begin{equation}\label{eq:fc-A-fiber-table}
\begin{array}{c|cc}
 \mu&\mathcal D_K^A(\mu)&\mathcal D_W^A(\mu)\\ \hline
 H&\{H\}&\operatorname{co}\{H,U,V\}\\
 U&\{U\}&\operatorname{co}\{U,H\}\\
 V&\{V\}&\operatorname{co}\{V,H\}
\end{array}.
\end{equation}
For example, $(H+U)/2$ belongs to the $A$ Wasserstein information fiber at future $H$ but is not itself a law in $\mathcal C_\mathrm{fc}$. In contrast, the $F$ fibers are full metric balls in $\mathcal P$, not convex hulls of future laws. In either experiment, common IID history $H$ is admissible for the three worlds $(K,H)$, $(W,U)$, and $(W,V)$, because $H\in\mathcal C_\mathrm{fc}$ and \eqref{eq:fc-metric-table} gives the required distances.

For completeness, convexifying the $A$ component set does \emph{not} amount to assuming the convexified law is an admissible historical marginal. If $Q=\sum_{j=1}^m\alpha_j\nu^{(j)}$ with each $\nu^{(j)}\in\mathcal C_\mathrm{fc}\cap B_d(\mu,\varepsilon)$, let an adversary independently draw each historical marginal $\nu_i$ with probabilities $\alpha_j$, and then draw $X_i\mid\nu_i\sim\nu_i$ independently. Every realized vector of marginals is deterministically admissible under $A$. Its unconditional sample law is exactly $Q^{\otimes n}$. Hence, for any measurable randomized policy $\pi_n$ and the induced unconditional output-price law $q_{n,Q}$ under $Q^{\otimes n}$,
\begin{equation}\label{eq:fc-A-random-adversary}
 \ell_\mu(q_{n,Q})
 =\mathbb E_{Q^{\otimes n}}\ell_\mu(\pi_n(\cdot\mid X_{1:n}))
 \le R_{d,n}^A(\pi).
\end{equation}
The inequality integrates the risk of each admissible deterministic component array and therefore remains within the original supremum, even for labeled or order-dependent policies. This establishes the relevant $A$ barycentric lower-bound interpretation, without asserting that the empirical average is sufficient for all product experiments. The exact finite-class frontier below uses the simpler common \emph{admissible} marginal $H$ for its matching lower bound.

\subsection{Exact constrained primal linear program}\label{app:fc-lp}

At common IID history $H$, project any randomized pricing distribution onto $\{t_1,t_2,t_3\}$. Write its probabilities in that order as $(x,y,z)\in\Delta_3$. The K future $H$ has regret $h=y/40+3z/200$, and the W futures $U,V$ have regrets
\begin{equation}\label{eq:fc-three-losses}
 h=\frac{y}{40}+\frac{3z}{200},\qquad
 u=\frac{11x}{100}+\frac{y}{100},\qquad
 v=\frac{y}{40}+\frac z2.
\end{equation}
Since $v\ge h$, the W future $H$ is redundant in this constrained game. Consequently the common-history linear program is
\begin{equation}\label{eq:fc-explicit-lp}
 \min_{(x,y,z)\in\Delta_3}\max\{u,v\}
 \quad\text{subject to}\quad h\le b.
\end{equation}
The following three feasible solutions cover the entire nonnegative budget domain:
\begin{equation}\label{eq:fc-piecewise-primal}
(x_b,y_b,z_b)=
\begin{cases}
\displaystyle\left(1-\frac{200b}{3},\ 0,\ \frac{200b}{3}\right),
 &0\le b\le\frac{33}{12200},\\[6pt]
\displaystyle\left(\frac{491}{535}-\frac{3880}{107}b,\quad
 \frac{4880}{107}b-\frac{66}{535},\quad
 \frac{22}{107}-\frac{1000}{107}b\right),
 &\frac{33}{12200}\le b\le\frac{11}{500},\\[6pt]
\displaystyle\left(\frac3{25},\ \frac{22}{25},\ 0\right),
 &b\ge\frac{11}{500}.
\end{cases}
\end{equation}
The entries are nonnegative and sum to one on their specified intervals. At the first breakpoint the two displayed policies agree and equal $(50/61,0,11/61)$; at the second they agree and equal $(3/25,22/25,0)$. For the first segment, $h=b$, $u=11/100-(22/3)b$ and $v=(100/3)b\le u$. For the middle segment, $h=b$ and
\[
 u=v=\frac{1067}{10700}-\frac{378}{107}b.
\]
For the last segment, $h=u=v=11/500\le b$. Thus all three primal candidates attain the right side of \eqref{eq:fc-piecewise}, and $h\le\max\{u,v\}$ at every budget. Substituting $b=1/50$ into the middle formula gives
\[
 (x_b,y_b,z_b)=\left(\frac{103}{535},\frac{422}{535},\frac{2}{107}\right),\qquad
 (h,u,v)=\left(\frac1{50},\frac{311}{10700},\frac{311}{10700}\right).
\]
Notice that the precise ordering of $x,y,z$ is $(t_1,t_2,t_3)$, whereas some original research notes use the reverse order for their first two coordinates.

\subsection{Matching all-price duals and global statistical lower bound}\label{app:fc-dual}

For nonnegative $\lambda_U,\lambda_V$ summing to one and a nonnegative K multiplier $m$, an all-price dual certificate consists of
\begin{equation}\label{eq:fc-dual-general}
 \lambda_U\ell_U(p)+\lambda_V\ell_V(p)+m\ell_H(p)\ge c
 \quad\text{for every }p\in[0,1].
\end{equation}
The following exact certificates are sufficient:
\begin{equation}\label{eq:fc-dual-certificates}
\begin{array}{c|cccc}
 \text{budget range}&\lambda_U&\lambda_V&m&c\\ \hline
 {\lbrack 0,33/12200\rbrack}&1&0&22/3&11/100\\
 {\lbrack 33/12200,11/500\rbrack}&97/107&10/107&378/107&1067/10700\\
 {\lbrack 11/500,\infty)}&1/5&4/5&0&11/500
\end{array}.
\end{equation}
To check every price, apply the componentwise regret projection established in \S\ref{app:fc-support}. On $(t_1,t_2,t_3)$, the weighted regret vectors for the three rows are, respectively,
\[
 \left(\frac{11}{100},\frac{29}{150},\frac{11}{100}\right),\quad
 \left(\frac{1067}{10700},\frac{1067}{10700},\frac{1067}{10700}\right),\quad
 \left(\frac{11}{500},\frac{11}{500},\frac25\right).
\] Because all multipliers are nonnegative, dominated prices cannot have lower weighted regret. This is an exact verification of \eqref{eq:fc-dual-general} on the \emph{entire} price interval, not just at the primal support.

Integrating \eqref{eq:fc-dual-general} against any randomized price law $q$ gives
\begin{equation}\label{eq:fc-dual-risk}
 \max\{\ell_U(q),\ell_V(q)\}\ge
 \lambda_U\ell_U(q)+\lambda_V\ell_V(q)
 \ge c-m\ell_H(q).
\end{equation}
If $\ell_H(q)\le b$, the dual lower bound is $c-mb$. The three lines $c-mb$ from \eqref{eq:fc-dual-certificates} coincide, respectively, with all three pieces achieved by \eqref{eq:fc-piecewise-primal}. Their intersections occur at the two stated breakpoints, including equality at both ends. This proves exact strong primal--dual equality for \eqref{eq:fc-explicit-lp}, without any unproved convexity or continuity assertion about an infinite-dimensional statistical game.

For the \emph{global statistical} lower bound, fix $n$ and an arbitrary policy kernel $\pi_n$. Let $q_{n,H}$ be its unconditional posted-price law when $X_i$ are IID from $H$. Since the identical sample law $H^{\otimes n}$ is allowed under the three worlds $(K,H)$, $(W,U)$, $(W,V)$ in both experiments,
\begin{equation}\label{eq:fc-lower-for-every-n}
 \ell_H(q_{n,H})\le R_{K,n}^E(\pi),\quad
 \max\{\ell_U(q_{n,H}),\ell_V(q_{n,H})\}\le R_{W,n}^E(\pi).
\end{equation}
Combining \eqref{eq:fc-dual-risk} and \eqref{eq:fc-lower-for-every-n} yields
\[
 R_{W,n}^E(\pi)\ge c-mR_{K,n}^E(\pi)\quad\text{for every }n.
\]
If $R_K^E(\pi)\le b$, then for every $\delta>0$ the K bound $R_{K,n}^E\le b+\delta$ holds for all sufficiently large $n$, by the definition of $\limsup$. Therefore $R_W^E(\pi)\ge c-m(b+\delta)$; let $\delta\downarrow0$. This proves the piecewise statistical lower bound simultaneously for $F$ and $A$. No interchange of $\sup$, expectation, and $\limsup$, or limit of pricing distributions, is used here. A negative K budget has no feasible policy, since regret is nonnegative.

\subsection{Uniformly valid non-IID statistical implementation}\label{app:fc-classifier}

Fix $b\ge0$ and let $q_b$ be the three-price law in \eqref{eq:fc-piecewise-primal}. Define the following deterministic measurable classification, followed by the indicated randomized price kernel. For the empirical historical measure $\widehat\nu_n=n^{-1}\sum_i\delta_{X_i}$, in the order listed,
\begin{equation}\label{eq:fc-global-classifier}
 \pi_{n,b}(\cdot\mid X_{1:n})=
 \begin{cases}
 \delta_{t_3},&d_K(\widehat\nu_n,U)<1/20,\\
 \delta_{t_1},&d_K(\widehat\nu_n,V)<1/20\ \text{and the first test fails},\\
 q_b,&\text{otherwise}.
 \end{cases}
\end{equation}
Both distance tests are Borel functions of the ordered observations (CDF suprema can be evaluated on a fixed countable determining set and the finitely many jump points). Strict inequalities and the stated order assign every equality case unambiguously. The final branch contains every unclassified observation vector; there is no assumption that the true array, its barycenter, or the realized empirical distribution belongs to $\mathcal C_\mathrm{fc}$.

For independent, not necessarily identically distributed $X_i\sim\nu_i$, put $\bar\nu_n=n^{-1}\sum_i\nu_i$ and $\xi_n=d_K(\widehat\nu_n,\bar\nu_n)$. Lemma~\ref{lem:fs-empirical} of the already established finite-sample appendix gives, uniformly over all independent arrays,
\[
 \mathbb E\xi_n\le4n^{-1/2}.
\]
Take $a_n=n^{-1/4}$ and $\mathsf G_n=\{\xi_n\le a_n\}$. Markov's inequality proves the uniform, distribution-free failure probability
\begin{equation}\label{eq:fc-uniform-good-event}
 \sup_{\nu_{1:n}\ \mathrm{independent}}
 \Pr(\mathsf G_n^c)\le4n^{-1/4}\longrightarrow0.
\end{equation}
In what follows take $n$ large enough that $a_n<1/25$. Under a $d$-admissible array, \eqref{eq:fc-average-radius} gives $d(\bar\nu_n,\mu)\le1/100$.

\paragraph{Separation, including the boundary.}
For any distribution $Q$ and any $0\le\rho<1/4$, the CDF intervals $I_1,I_2$ give the \emph{two} symmetric inequalities
\begin{equation}\label{eq:fc-separation}
 \begin{split}
 d_K(Q,U)\le\rho&\ \Longrightarrow\
 d_W(Q,V)\ge\frac1{50}-\frac4{75}\rho,\\
 d_K(Q,V)\le\rho&\ \Longrightarrow\
 d_W(Q,U)\ge\frac1{50}-\frac4{75}\rho.
 \end{split}
\end{equation}
Indeed, on the respective intervals the CDF discrepancy from the opposing law is at least $1/4-\rho$ and $3/4-\rho$; integrating their positive lower bounds gives
$ (1/25)(1/4-\rho)+(1/75)(3/4-\rho)=1/50-4\rho/75$.
If the first test in \eqref{eq:fc-global-classifier} succeeds on $\mathsf G_n$, then $d_K(\bar\nu_n,U)<1/20+a_n<9/100$. Equation~\eqref{eq:fc-separation} gives
\[
 d_W(\bar\nu_n,V)>\frac1{50}-\frac4{75}\frac9{100}
 =\frac{19}{1250}>\frac1{100}.
\]
The symmetric conclusion applies to the second test and W future $U$. Therefore a Wasserstein-admissible future cannot be $V$ on the U branch, or $U$ on the V branch, on the good event. The strict inequality persists even if the true average lies \emph{exactly on} its closed-radius boundary: the empirical threshold has enough fixed separation slack. No assertion of strict separation by the triangle inequality at $d_W(U,V)=2\varepsilon$ is used.

\paragraph{Kolmogorov safety.}
For a K-admissible future $U$, $d_K(\widehat\nu_n,U)\le1/100+a_n<1/20$ on $\mathsf G_n$, so the U branch posts $t_3$ and has zero U regret. For future $V$, the U branch is impossible because $d_K(U,V)=3/4$ while $d_K(\widehat\nu_n,V)\le1/100+a_n$, and the V branch posts $t_1$ and has zero V regret. For future $H$, both early tests fail by $d_K(H,U)=1/4$, $d_K(H,V)=3/4$, and the bound $1/100+a_n$; the central branch has $\ell_H(q_b)\le b$. Thus, uniformly over every K-admissible array in either experiment, conditional expected regret on $\mathsf G_n$ is at most $b$.

\paragraph{Wasserstein performance.}
For future $U$, the V branch is ruled out by \eqref{eq:fc-separation}; the remaining U and central branches have U regrets $0$ and $u(q_b)\le G_\mathrm{fc}(b)$, respectively, where $G_\mathrm{fc}(b)$ denotes the right side of \eqref{eq:fc-piecewise}. For future $V$, the U branch is excluded; its V and central branch regrets are $0$ and $v(q_b)\le G_\mathrm{fc}(b)$. Future $H$ may enter any branch, but the corresponding H regrets are $3/200$, $0$, and $h(q_b)$. These are all at most $G_\mathrm{fc}(b)$, since $G_\mathrm{fc}(b)\ge11/500>3/200$ and $h(q_b)\le G_\mathrm{fc}(b)$ by \eqref{eq:fc-piecewise-primal}. On $\mathsf G_n^c$ the regret is bounded by $1$. Thus for all sufficiently large $n$,
\begin{equation}\label{eq:fc-upper-risk-n}
 R_{K,n}^E(\pi_{\cdot,b})\le b+4n^{-1/4},\qquad
 R_{W,n}^E(\pi_{\cdot,b})\le G_\mathrm{fc}(b)+4n^{-1/4}.
\end{equation}
The inequalities hold for every deterministic admissible component array before taking the environmental supremum and do not require identical marginals. Taking $\limsup_{n\to\infty}$ at the fixed radius proves global K feasibility and the global W upper bound for \emph{each} $b\ge0$, including $b=0$, $b=33/12200$, and $b=11/500$. Combined with the lower bound, this completes \eqref{eq:fc-piecewise} for both $F$ and $A$.

\subsection{Exact W oracle and its access budget}\label{app:fc-access}

The last row of \eqref{eq:fc-dual-certificates}, with $m=0$, gives a W-only lower bound at every $n$ in both experiments:
\[
 R_{W,n}^E(\pi)\ge\frac{11}{500}.
\]
The policy \eqref{eq:fc-global-classifier} at $b=11/500$ has central price mixture
\begin{equation}\label{eq:fc-unique-oracle-mixture}
 q_W=\frac3{25}\delta_{t_1}+\frac{22}{25}\delta_{t_2},
 \qquad \ell_H(q_W)=\ell_U(q_W)=\ell_V(q_W)=\frac{11}{500},
\end{equation}
so $W_E^\star=11/500$ and $b_E^W\le11/500$.

To prove the reverse bound on the \emph{least K risk of a W-oracle policy}, fix any policy sequence with $R_W^E(\pi)=11/500$ and again let $q_{n,H}$ be its unconditional price distribution under IID $H$. Project each $q_{n,H}$ to a three-price law $\widetilde q_n$ using the componentwise domination map from \S\ref{app:fc-support}. We then have
\[
 \ell_\mu(\widetilde q_n)\le\ell_\mu(q_{n,H})\quad(\mu=H,U,V),
 \qquad\limsup_n\ell_U(\widetilde q_n),\ \limsup_n\ell_V(\widetilde q_n)\le\frac{11}{500}.
\]
Any accumulation point $(x,y,z)$ of the probability vectors of $\widetilde q_n$ satisfies $u\le11/500$ and $v\le11/500$. The W-only weighted dual has values $11/500,11/500,2/5$ on the three prices, so
\[
 \frac15u+\frac45v=\frac{11}{500}+\left(\frac25-\frac{11}{500}\right)z
 \le\frac{11}{500}.
\]
It follows that $z=0$. With $x+y=1$, the inequalities
\[
 u=\frac1{100}+\frac{x}{10}\le\frac{11}{500},\qquad
 v=\frac{1-x}{40}\le\frac{11}{500}
\]
force $x=3/25$ and $y=22/25$. Thus every accumulation point equals the unique vector in \eqref{eq:fc-unique-oracle-mixture}; compactness of $\Delta_3$ implies $\ell_H(\widetilde q_n)\to11/500$. By domination,
\[
 \liminf_n\ell_H(q_{n,H})\ge\frac{11}{500},\qquad
 R_K^E(\pi)\ge\limsup_n\ell_H(q_{n,H})\ge\frac{11}{500}.
\]
This conclusion requires no optimizer or weak-continuity assumption for general pricing laws. Since the constructed global policy attains W-oracle risk with K risk at most $11/500$, it proves $b_E^W=11/500$. Finally,
\[
 \frac{311}{10700}-\frac{11}{500}=\frac{189}{26750}>0
\]
proves \eqref{eq:fc-gap}.

\subsection{Why one K/W pair is insufficient}\label{app:fc-pairwise}

For context, define the ordered one-K-future/one-W-future certificate at $b=1/50$ as the supremum, over pairs $(\mu_K,\mu_W)$ having at least one common admissible historical marginal and over the implied two-world constrained pricing problem, of
\[
 \min_{q:\,\ell_{\mu_K}(q)\le1/50}\ell_{\mu_W}(q).
\]
The condition on a common marginal is $d_K(Q,\mu_K)\le1/100$ and $d_W(Q,\mu_W)\le1/100$ in $F$, with $Q\in\mathcal C_\mathrm{fc}$ in $A$. It is sufficient for excluding pairs in both experiments: by \eqref{eq:fc-separation} with $\rho=1/100$, the ordered pairs $(U,V)$ and $(V,U)$ are impossible even in $F$. Conversely, each other ordered pair has a common historical witness in $\mathcal C_\mathrm{fc}$ (take the K future itself if the laws are distinct, or their identical law).

Except for $(\mu_K,\mu_W)=(U,H)$, every remaining pair has a price with zero W regret and K regret at most $1/50$: choose $t_3$ for $(H,U)$, $t_1$ for $(H,V)$ and $(V,H)$, and a respective optimal price for identical pairs. For $(U,H)$, the only relevant optimum mixes $t_1$ and $t_3$, with respective probabilities $2/11$ and $9/11$: its U regret is $(11/100)(2/11)=1/50$, and H regret is $(3/200)(9/11)=27/2200$. Optimality holds globally because
\[
 \ell_H(p)+\frac3{22}\ell_U(p)\ge\frac3{200}
 \quad\text{for every }p\in[0,1],
\]
which is verified at the three undominated prices and extended by support domination. Therefore every U-safe rule has H regret at least $3/200-(3/22)(1/50)=27/2200$. The pair is genuinely attainable in $A$ as well as $F$ by homogeneous history $U$. Hence the best such pairwise certificate is precisely
\begin{equation}\label{eq:fc-pairwise-value}
 S_2^F=S_2^A=\frac{27}{2200}
 <\frac{311}{10700}=G_E(1/50).
\end{equation}
Moreover $27/2200<11/500=W_E^\star$, so $[S_2^E-W_E^\star]_+=0$ even though the global excess in \eqref{eq:fc-gap} is strictly positive. The two positive W multipliers and the positive K multiplier in the middle dual row of \eqref{eq:fc-dual-certificates} capture the simultaneous multi-future obstruction absent from any isolated pair.

\section{Proof of Comparable Unequal-Radius Robustness}\label{app:ur-proof}

We prove Proposition~\ref{prop:ur-frontier} under the observation and risk definitions of Section~\ref{sec:model}. Fix $0<\alpha<1$, $0<c\le C<\infty$, and $\mathcal C=\mathcal H_{\alpha,5}$. For every fixed pair $(\varepsilon_K,\varepsilon_W)$ the policy is one sequence of measurable stochastic kernels; it is not informed which metric environment occurs. Write $q=\alpha/(1+\alpha)\in(0,1/2)$ and use $\Phi_\alpha$ from \eqref{eq:ur-frontier}. All distributions below are supported on $[0,1]$, with purchase when the valuation \emph{equals} the price.

\subsection{Monotonicity and uniform comparison of radii}\label{app:ur-comparison}

\begin{lemma}[Separate-radius monotonicity]\label{lem:ur-monotonicity}
For $E\in\{F,A\}$ and fixed $\mathcal C$, $G_E(b;\mathcal C,r_K,r_W)$ is nondecreasing in each of $r_K$ and $r_W$, for any fixed budget $b$. In particular, if $r_-:=\min\{\varepsilon_K,\varepsilon_W\}$, then
\begin{equation}\label{eq:ur-lower-comparison}
 G_E(b;\mathcal C,\varepsilon_K,\varepsilon_W)
 \ge G_E(b;\mathcal C,r_-,r_-).
\end{equation}
Furthermore,
\begin{align}
 \varepsilon_K\ge\varepsilon_W\quad&\Longrightarrow\quad
 G_E(b;\mathcal C,\varepsilon_K,\varepsilon_W)
 \le G_E(b;\mathcal C,\varepsilon_K,\varepsilon_K),\label{eq:ur-upper-comparison-k}\\
 \varepsilon_K<\varepsilon_W\quad&\Longrightarrow\quad
 G_E(b;\mathcal C,\varepsilon_K,\varepsilon_W)
 \le G_E(b;\mathcal C,\varepsilon_W,\varepsilon_W).\label{eq:ur-upper-comparison-w}
\end{align}
\end{lemma}
\begin{proof}
For $d=K,W$, increasing radius $r$ enlarges $\mathfrak E_{d,n}^{E}(\mathcal C,r)$ at every $n$ in \eqref{eq:model-environments}. The expected regret of any fixed kernel is unchanged at an individual environment, hence $R_{d,n}^E(\pi;\mathcal C,r)$ and then its $\limsup_n$ are nondecreasing in $r$. Increasing $r_K$ can only shrink the feasible policy-sequence set in \eqref{eq:ur-definition}; increasing $r_W$ can only raise the objective of each feasible sequence. Both operations weakly raise the infimum, including if the feasible set is empty. Applying these two monotonicities separately gives \eqref{eq:ur-lower-comparison}--\eqref{eq:ur-upper-comparison-w}. No inclusion \emph{between} a K ball and a W ball at unequal radii is used.
\end{proof}

\begin{lemma}[Budget-uniform radius equivalence]\label{lem:ur-radius-equivalence}
Set $m=\min\{1,c\}\in(0,1]$ and $M=\max\{1,C\}\ge1$. For every $r>0$, $\lambda\in[m,M]$, and $b>0$,
\begin{equation}\label{eq:ur-scale-comparison}
 m\,\Phi_\alpha(r,b)\le\Phi_\alpha(\lambda r,b)
 \le M\,\Phi_\alpha(r,b).
\end{equation}
Thus \eqref{eq:ur-scale-comparison} applies both to $\lambda=r_-/\varepsilon_W$ and to $\lambda=\varepsilon_K/\varepsilon_W$, uniformly in $b$, including budgets that vary with both radii.
\end{lemma}
\begin{proof}
Each of the three positive entries defining $\Phi_\alpha(\lambda r,b)$ is its counterpart at $r$ multiplied by $\lambda^{1/2}$, $\lambda^q$, or $\lambda$, respectively. Since $0<q<1/2<1$ and $m\le\lambda\le M$, all three multipliers lie in $[m,M]$. The maps $\min$ and $\max$ preserve componentwise inequalities, giving \eqref{eq:ur-scale-comparison}. The constants are deterministic and do not involve $b$. The ratio hypotheses imply $r_-/\varepsilon_W\in[m,M]$ and $\varepsilon_K/\varepsilon_W\in[m,M]$.
\end{proof}

\subsection{Future-H\"older interpolation with unrestricted history}\label{app:ur-interpolation}

The subtle upper-bound case requires an explicit estimate with regularity on the \emph{future}, not on the historical marginal. We give a coupling proof independently of the integrated-CDF argument in Lemma~\ref{lem:holder-interpolation}.

\begin{lemma}[One-sided regularity suffices]\label{lem:ur-future-interpolation}
Let $\mu\in\mathcal H_{\alpha,5}$, and let $\rho$ be \emph{any} Borel probability law on $[0,1]$, including one with atoms. For every $h>0$,
\begin{equation}\label{eq:ur-coupling-interpolation}
 d_K(\rho,\mu)\le5h^\alpha+\frac{W_1(\rho,\mu)}h.
\end{equation}
Consequently, for $0<\delta\le1$, $W_1(\rho,\mu)\le\delta$ implies
\begin{equation}\label{eq:ur-interpolated-rate}
 d_K(\rho,\mu)\le6\delta^q,
 \qquad\sup_{p\in[0,1]}|r_\rho(p)-r_\mu(p)|\le6\delta^q.
\end{equation}
The bound includes prices at historical atoms and the $V\ge p$ purchase convention.
\end{lemma}
\begin{proof}
The interval-mass restriction implies $\mu(\{x\})=0$ for every $x$. Let $(X,Y)$ be an optimal coupling with $X\sim\rho$, $Y\sim\mu$; existence follows from compactness. For $x\in[0,1]$,
\[
 F_\rho(x)-F_\mu(x)
 =\mathbb P\{X\le x<Y\}-\mathbb P\{Y\le x<X\}.
\]
On $\{|X-Y|\le h\}$ the first event forces $Y\in(x,x+h]$, whose $\mu$-probability is at most $5h^\alpha$. The remaining probability is at most $\mathbb P\{|X-Y|>h\}\le W_1(\rho,\mu)/h$. Similarly, on $\{|X-Y|\le h\}$ the second event forces $Y\in[x-h,x]$, again of $\mu$-mass at most $5h^\alpha$. Thus both signs of $F_\rho(x)-F_\mu(x)$ are bounded by the right side of \eqref{eq:ur-coupling-interpolation}, including at endpoints (intervals are intersected with $[0,1]$). Take the supremum over $x$. If $W_1\le\delta\le1$, choose $h=\delta^{1/(1+\alpha)}$ to get $5\delta^q+\delta^q=6\delta^q$.

For $p>0$, $\rho([p,1])=1-F_\rho(p-)$ and likewise for $\mu$. The uniform CDF estimate passes to the \emph{left} limits, and multiplying the tail difference by $p\le1$ gives the revenue estimate. At $p=0$ both revenues are zero. In particular, the atom of $\rho$ \emph{at} $p$ is included in $\rho([p,1])$; no atomless or H\"older assumption on $\rho$ was made.
\end{proof}

\subsection{An implementable policy at the closed K budget}\label{app:ur-saa}

For independent, not necessarily identically distributed historical valuations $X_i\sim\nu_i$, define $\bar\nu_n=n^{-1}\sum_{i=1}^n\nu_i$, $\widehat\nu_n=n^{-1}\sum_{i=1}^n\delta_{X_i}$, and
\[
 \widehat r_n(p)=r_{\widehat\nu_n}(p),\qquad
 \eta_n=\sup_{p\in[0,1]}|\widehat r_n(p)-r_{\bar\nu_n}(p)|.
\]
Choose $\widehat p_n$ as the smallest maximizer of $\widehat r_n$ among $\{0,X_1,\ldots,X_n\}$ and let $\pi_n^{\rm SAA}(\cdot\mid X_{1:n})=\delta_{\widehat p_n}$. Between distinct observed positive valuations empirical revenue increases linearly, with any downward jump occurring \emph{after} an observed value. Hence the chosen price maximizes $\widehat r_n$ over all $[0,1]$, including ties and zero observations. This is a Borel function of the full sample (a finite maximization with a fixed tie rule), so $\pi^{\rm SAA}$ is a genuine deterministic stochastic-kernel sequence and uses neither unknown $\mu$ nor the true history laws.

\begin{lemma}[Simultaneous separate-radius SAA risks]\label{lem:ur-saa}
For any $\varepsilon_K>0$ and $0<\varepsilon_W\le1$, the same $\pi^{\rm SAA}$ satisfies, for each $E=F,A$ and every $n\ge1$,
\begin{align}
 R_{K,n}^E(\pi^{\rm SAA};\mathcal H_{\alpha,5},\varepsilon_K)
 &\le2\varepsilon_K+\frac8{\sqrt n},\label{eq:ur-saa-finite-k}\\
 R_{W,n}^E(\pi^{\rm SAA};\mathcal H_{\alpha,5},\varepsilon_W)
 &\le12\varepsilon_W^q+\frac8{\sqrt n}.\label{eq:ur-saa-finite-w}
\end{align}
Consequently,
\begin{equation}\label{eq:ur-saa-asymptotic}
 R_K^E(\pi^{\rm SAA};\mathcal H_{\alpha,5},\varepsilon_K)
 \le2\varepsilon_K,\qquad
 R_W^E(\pi^{\rm SAA};\mathcal H_{\alpha,5},\varepsilon_W)
 \le12\varepsilon_W^q.
\end{equation}
In particular, feasibility holds at the \emph{closed} endpoint $b=2\varepsilon_K$ in the stipulated asymptotic-risk sense; no finite-$n$ zero-slack certificate is claimed.
\end{lemma}
\begin{proof}
The CDF of the historical barycenter is the arithmetic mean of the component CDFs. Thus, for either metric $d\in\{K,W\}$ and any $d$-admissible independent array,
\begin{equation}\label{eq:ur-barycenter}
 d(\bar\nu_n,\mu)
 \le\frac1n\sum_{i=1}^n d(\nu_i,\mu)
 \le\varepsilon_d.
\end{equation}
For $d_K$ this follows by the triangle inequality under the supremum norm; for $W_1$ it follows by integrating $|F_{\bar\nu_n}-F_\mu|$ or by mixing optimal couplings. The latter does \emph{not} require any $\nu_i$ or $\bar\nu_n$ to be H\"older, or $\bar\nu_n$ to belong to $\mathcal H_{\alpha,5}$.

By the non-IID empirical-process proof in Lemma~\ref{lem:fs-empirical} (also Lemma~\ref{lem:holder-noniid}),
\begin{equation}\label{eq:ur-noniid-deviation}
 \sup_{\nu_{1:n}\in\mathcal P^n}\mathbb E_{\nu_{1:n}}\eta_n\le\frac4{\sqrt n}.
\end{equation}
For completeness, introduce a separate independent ghost valuation with law $\nu_i$ for every coordinate and symmetrize with independent Rademacher signs. Conditional on the realized observations, sets $\{i:X_i\ge p\}$ are suffixes of their valuation ordering (ties only remove candidate suffixes), while the factor $p$ is at most one. The signed revenue-process supremum is bounded by the maximal absolute partial sum of $n$ Rademacher signs; Doob's $L^2$ inequality bounds its expectation by $2\sqrt n$. The symmetrization factor two gives \eqref{eq:ur-noniid-deviation}. Identical distributions, continuity, and assumptions on historical atoms are unnecessary.

For every realized sample, empirical optimality yields
\begin{equation}\label{eq:ur-empirical-optimality}
 r_{\bar\nu_n}(\widehat p_n)\ge r_{\bar\nu_n}^\star-2\eta_n.
\end{equation}
Indeed, compare $\widehat r_n$ to $r_{\bar\nu_n}$ at a population maximizing price and at $\widehat p_n$. A maximizing price exists: $p\mapsto p\rho([p,1])$ is upper semicontinuous on $[0,1]$ for any $\rho$, because its value includes purchases at equality. Let $D=\sup_p|r_\mu(p)-r_{\bar\nu_n}(p)|$. The population optimal values differ by at most $D$, so the \emph{samplewise} regret satisfies
\begin{equation}\label{eq:ur-saa-comparison}
 r_\mu^\star-r_\mu(\widehat p_n)
 \le2D+2\eta_n.
\end{equation}
In a K-admissible environment, \eqref{eq:ur-barycenter} and the revenue comparison \eqref{eq:holder-k-revenue} give $D\le\varepsilon_K$. In a W-admissible environment, \eqref{eq:ur-barycenter} and Lemma~\ref{lem:ur-future-interpolation}, applied with $\rho=\bar\nu_n$ and the \emph{future} $\mu\in\mathcal H_{\alpha,5}$, give $D\le6\varepsilon_W^q$. Take independent-sample expectation in \eqref{eq:ur-saa-comparison}, then the separate environment suprema for $K$ and $W$, and apply \eqref{eq:ur-noniid-deviation}. This proves \eqref{eq:ur-saa-finite-k}--\eqref{eq:ur-saa-finite-w} uniformly in the admissible arrays for \emph{each} experiment. Finally take $\limsup_{n\to\infty}$ at the fixed radii to obtain \eqref{eq:ur-saa-asymptotic}. Only after that operation may a small-radius limit be considered.
\end{proof}

\subsection{All budget intervals and the endpoint}\label{app:ur-assembly}

\begin{proof}[Proof of Proposition~\ref{prop:ur-frontier}]
The approved uniform equal-radius Theorem~\ref{thm:holder-frontier} supplies constants $a_\alpha,A_\alpha>0$ and $r_0(\alpha)>0$ such that, for $E=F,A$, all $0<r\le r_0$ and all $b\ge2r$,
\begin{equation}\label{eq:ur-equal-input}
 a_\alpha\Phi_\alpha(r,b)
 \le G_E(b;\mathcal H_{\alpha,5},r,r)
 \le A_\alpha\Phi_\alpha(r,b).
\end{equation}
Choose $\varepsilon_0(\alpha,c,C)>0$ sufficiently small that $M\varepsilon_0\le r_0$, $\varepsilon_0\le1$, and $\varepsilon_0^q\le1/2$, where $m,M$ are as in Lemma~\ref{lem:ur-radius-equivalence}. Fix any allowable $\varepsilon_W,\varepsilon_K$ and $b$, including a choice that changes arbitrarily with $\varepsilon_W$.

\emph{Lower bound.} Let $r_-=\min\{\varepsilon_K,\varepsilon_W\}$. Because $b\ge2\varepsilon_K\ge2r_-$, the certified result \eqref{eq:ur-equal-input} applies at $r_-$. Lemmas~\ref{lem:ur-monotonicity} and \ref{lem:ur-radius-equivalence} give
\begin{equation}\label{eq:ur-final-lower}
 G_E(b;\mathcal H_{\alpha,5},\varepsilon_K,\varepsilon_W)
 \ge G_E(b;\mathcal H_{\alpha,5},r_-,r_-)
 \ge a_\alpha\Phi_\alpha(r_-,b)
 \ge a_\alpha m\Phi_\alpha(\varepsilon_W,b).
\end{equation}
The same inequality holds in both $F$ and $A$ and requires no change to the approved equal-radius witnesses.

\emph{Upper bound I: $\varepsilon_K\ge\varepsilon_W$.} Here $b\ge2\varepsilon_K$, so the equal-radius upper bound at $r=\varepsilon_K$ is available. By \eqref{eq:ur-upper-comparison-k} and \eqref{eq:ur-scale-comparison},
\begin{equation}\label{eq:ur-final-upper-k}
 G_E(b;\mathcal H_{\alpha,5},\varepsilon_K,\varepsilon_W)
 \le G_E(b;\mathcal H_{\alpha,5},\varepsilon_K,\varepsilon_K)
 \le A_\alpha M\Phi_\alpha(\varepsilon_W,b).
\end{equation}

\emph{Upper bound II: $\varepsilon_K<\varepsilon_W$ and $b\ge2\varepsilon_W$.} This budget is eligible for the equal-radius frontier at $r=\varepsilon_W$. Equation~\eqref{eq:ur-upper-comparison-w} gives
\begin{equation}\label{eq:ur-final-upper-w}
 G_E(b;\mathcal H_{\alpha,5},\varepsilon_K,\varepsilon_W)
 \le G_E(b;\mathcal H_{\alpha,5},\varepsilon_W,\varepsilon_W)
 \le A_\alpha\Phi_\alpha(\varepsilon_W,b).
\end{equation}

\emph{Upper bound III: $\varepsilon_K<\varepsilon_W$ and $2\varepsilon_K\le b<2\varepsilon_W$.} This interval is \emph{not} covered by an equal-$\varepsilon_W$ comparison, because that comparison would demand the stronger safety budget $b\ge2\varepsilon_W$. Lemma~\ref{lem:ur-saa}, however, supplies one observation-based policy with $R_K^E\le2\varepsilon_K\le b$ and $R_W^E\le12\varepsilon_W^q$ for either experiment. Moreover $b<2\varepsilon_W$ implies $\varepsilon_W/b>1/2\ge\varepsilon_W^q$; since $q<1/2$ and $\varepsilon_W\le1$, $\sqrt{\varepsilon_W}\le\varepsilon_W^q$. Hence the \emph{exact expression for the rate envelope} throughout this entire missing interval is
\begin{equation}\label{eq:ur-low-budget-envelope}
 \Phi_\alpha(\varepsilon_W,b)=\varepsilon_W^q,
\end{equation}
and consequently
\begin{equation}\label{eq:ur-final-upper-low}
 G_E(b;\mathcal H_{\alpha,5},\varepsilon_K,\varepsilon_W)
 \le12\Phi_\alpha(\varepsilon_W,b).
\end{equation}
In particular, at $b=2\varepsilon_K$ the finite-sample $8/\sqrt n$ safety excess in \eqref{eq:ur-saa-finite-k} disappears under the \emph{prescribed} $\limsup_n$, so the endpoint is included without a budget perturbation or an exchange of limits. At $b=2\varepsilon_W$ the second case applies, so no budget gap remains.

Combining \eqref{eq:ur-final-lower} with the three upper bounds gives Proposition~\ref{prop:ur-frontier}, e.g., with $k_{\alpha,c,C}=a_\alpha m$ and $K_{\alpha,c,C}=\max\{A_\alpha M,12\}$. These constants apply to every eligible $(\varepsilon_K,\varepsilon_W,b)$ jointly; allowing arbitrary budget sequences cannot change them. The rate interpretation follows from $q<1/2$ and the two scale comparisons in $\Phi_\alpha$. Nothing in the argument identifies exact unequal-radius minimax coefficients, finite-radius regime kinks, or oracle-access thresholds, nor covers radius ratios escaping $[c,C]$.
\end{proof}

\bibliographystyle{plainnat}
\bibliography{references}
\end{document}